# Photonic Neural Networks and Optics-informed Deep Learning Fundamentals

Apostolos Tsakyridis[1*†], Miltiadis Moralis-Pegios[1†], George Giamougiannis[1], Manos Kirtas[1], Nikolaos Passalis[1], Anastasios Tefas[1], Nikos Pleros[1]

[1]Department of Informatics, Aristotle University of Thessaloniki, 54124, Thessaloniki, Greece

[†] The authors contributed equally to this work.

*atsakyrid@csd.auth.gr

**Abstract**

The recent explosive compute growth, mainly fueled by the boost of artificial intelligence (AI) and deep neural networks (DNNs), is currently instigating the demand for a novel computing paradigm that can overcome the insurmountable barriers imposed by conventional electronic computing architectures. Photonic neural networks (PNNs) implemented on silicon integration platforms stand out as a promising candidate to endow neural network (NN) hardware, offering the potential for energy efficient and ultra-fast computations through the utilization of the unique primitives of photonics i.e. energy efficiency, THz bandwidth and low-latency. Thus far, several demonstrations have revealed the huge potential of PNNs in performing both linear and non-linear NN operations at unparalleled speed and energy consumption metrics. Transforming this potential into a tangible reality for deep learning (DL) applications requires, however, a deep understanding of the basic PNN principles, requirements and challenges across all constituent architectural, technological and training aspects. In this tutorial, we, initially, review the principles of DNNs along with their fundamental building blocks, analyzing also the key mathematical operations needed for their computation in a photonic hardware. Then, we investigate, through an intuitive mathematical analysis, the interdependence of bit precision and energy efficiency in analog photonic circuitry, discussing the opportunities and challenges of PNNs. Followingly, a performance overview of PNN architectures, weight technologies and activation functions is presented, summarizing their impact in speed, scalability and power consumption. Finally, we provide an holistic overview of the optics-informed NN training framework that incorporates the physical properties of photonic building blocks into the training process in order to improve the NN classification accuracy and effectively elevate neuromorphic photonic hardware into high-performance DL computational settings.

1. **Introduction**

During the past decade, the relentless expansion of artificial intelligence (AI) through deep neural networks (DNNs), has been driving the need for high-performance computing and time-of-flight data processing. Conventional digital computing units, that are based on the well-known Von-Neumann architecture [1] and inherently rely on serialized data processing, have faced daunting challenges towards undertaking the execution of emerging DNN datasets. Von-Neumann architectures comprise a centralized processing unit (CPU), which is responsible for executing all operations (arithmetic, logic and controlling) dictated by the program's instructions, and a separate random-access memory (RAM) unit that stores all necessary data and instructions. The communication between CPU and memory is realized via a shared bus that is used to transfer all data between them, implying that they cannot be accessed simultaneously. This leads to the well-known Von-Neumann bottleneck [2], where the processor remains idle for a certain amount of time during memory data access. On top of that, the need for moving data between CPU and memory (via the bus) requires charging/discharging of metal wires, limiting in this way both the bandwidth and the energy efficiency due to Joule heating and capacitance [3], respectively.

There have been numerous demonstrations towards overcoming these effects, including, among others, caching, multithreading and new RAM architectures and technologies (e.g. ferroelectric RAMs [4] and optical RAMs [5]-[9]), with the ultimate target being the energy efficient and high-speed CPU-memory data movement. None of these solutions seems, however, to be capable of coping with the computational and energy demands of DNNs, revealing a need for shifting towards specialized computing hardware architectures. In this endeavour, highly parallelized accelerators have been developed, including graphic processing units (GPUs), application specific integrated circuits (ASICs) and field programmable gate arrays (FPGAs), with GPUs and ASICs being, until now, the dominant hardware computing engines for DNN implementations. Specifically, GPUs leverage their hundreds of cores towards accelerating the matrix multiplication operations of DNNs which are the most time- and power- consuming computations [10]. Moreover, they have dedicated non-uniform memory access architectures (e.g., Video RAMs) that are i) programmable, meaning that the stored data can be selectively accessed or deleted, ii) faster than CPU counterparts iii) located very close to their cores, reducing in this way the distance between computing and data. Yet, despite GPU's unrivalled parallelization ability that ushers in exceptional computational throughput, the need for data movement still remains and sets a fundamental limit in both speed and energy efficiency.

Towards totally eradicating the constraints of data movement, recent developments on analog computing through memristive crossbar arrays [11]-[13] follow an alternative approach, called in-memory computing. This scheme allows for certain DNN computational tasks (e.g. weighting) to be performed within the memory cell itself, seamlessly supporting multiplication operations without requiring any data transfer [14]. The recent 64-core analog-in-memory compute (AiMC) research prototype of IBM [15] and the commercial entry of Mythic's AiMC engine have validated the energy benefits that can originate from in-memory computing compared to Von-Neumann architectures. These implementations employ computational memory devices, including resistive RAMs (RRAMs), phase change materials (PCMs) etc., where the application of a voltage results in a change of the material's property, achieving in this way both data storing and computing. However, issues related to memory instability and finite resistance of the crossbar wires may lead to computational errors and crossbar size limitations, respectively, making it hard to reach the computational throughput and parallelization level of GPUs [12], [14]. Similar to in-memory computing, neuromorphic computing comprises an alternative non-Von-Neumann architecture that is inspired by the structure and function of the human brain, meaning that both memory and computing are governed by artificial neurons and synapses. Neuromorphic chips mostly employ spiking neural networks (SNNs) to emulate the behavior of biological neurons, which communicate through discrete electrical pulses called spikes.

SNNs can process spatiotemporal information more efficiently and accurately than conventional neural networks [16] as they respond to changes in the input data in real time. Additionally, they rely on asynchronous communication and event-driven computations, where, typically, only a small portion of the entire system is active at any given time while the rest is idle, resulting to low-power operation [17]. However, neuromorphic computing is not currently being used in real-word applications and there are still a wide variety of challenges in both algorithmic and application development [18], that need to be addressed towards outperforming conventional deep learning approaches. At the same time, the underlying electronic hardware in analog compute engines continues to rely heavily on complementary metal oxide semiconductor (CMOS) electronic transistors and interconnects, whose speed and energy efficiency are dictated by their size. Taking into account that transistor scaling has slowed down during the last decade, since we are approaching its fundamental physical size limits [19], there is no significant performance margin left to be gained. In parallel, the requirement for multiple connected neurons yields increased interconnect lengths in analog in-memory computing schemes that finally result to low line-rate operation in order to avoid an increased energy consumption. All this indicates that a radical departure from traditional electronic computing systems towards a novel computational hardware technology has to be realized in order to be able to fully reap the benefits of the architectural shift towards non-Von-Neumann layouts.

Along this direction, integrated photonics emerged as a promising candidate for the hardware implementation of DNNs; the analog nature of light is inherently compatible with analog compute principles, while low-energy and high-bandwidth connectivity is the natural advantage of optical wires. On top of that, photonics can offer multiple degrees of freedom such as wavelength, phase, mode and polarization, being suitable for parallelizing data processing through multiplexing techniques [20], [21] that have been traditionally employed in optical communication systems for transferring information at enormous datarates (>Tb/s). The constantly growing deployment of optical interconnects and their rapid penetration to smaller network segments has been also the driving force for the impressive advances witnessed in photonic integration and, particularly, in silicon photonics; silicon PICs with thousands of photonic components can be fabricated in a single die nowadays [22], forming a highly promising technology landscape for optical information processing tasks at chip-scale. Nevertheless, compared with electronic systems that host billions of transistors, thousands of photonic components may not be sufficient to build a vast universal hardware engine for generic applications. Yet, the constant progress in the field of integrated optics coupled with the rapid advances in fabrication and packaging can eventually shape new horizons in this field. This has raised expectations for an integrated photonic neural network (PNN) platform that can cope with the massively growing computational needs of Deep Learning (DL) engines, where computational capacity requirements double every 4-6 months [23]. In this realm, several PNN demonstrations have been proposed [24]-[41], employing light both for data transfer and computational functions and shaping a new roadmap for orders of magnitude higher computational and energy efficiencies than conventional electronic counterparts. At the same time, they highlighted a number of remaining challenges that have to be addressed at technology, architecture and training level, designating a bidirectional interactive relationship between hardware and software: the photonic hardware substrate has to comply with existing DL models and architectures, but at the same time the DL training algorithms have to adapt to the idiosyncrasy of the photonic hardware. Integrated neuromorphic photonic hardware extends along a pool of architectural and technology options, the main target being the deployment of highly scalable and energy efficient setups that are compatible with conventional DL training models and suitable to safeguard high accuracy performance. In parallel, the use of light in all its basic computational blocks brings inevitably a number of new physical and mathematical quantities in NN layouts [41], [42], such as noise and multiplication between "noisy" matrices, as well as mathematical expressions for non-typical activation responses, which are not encountered in conventional DL training models employed in the digital world. This calls for an optics-informed DL training model library; the term "optics-informed" has been recently coined by I. Roumpos et. al., [42] in order to describe the hardware-aware characteristics of DL training models and declare their alignment along the nature of optical hardware, since it takes into account the idiosyncrasy of light and photonic technology. However, despite the advances pursued in both the hardware and software segments, the complexity of photonic processing is still far behind electronics with respect to both their algorithmic and their hardware capabilities. Hence, the field of PNNs does not currently proceed along the mission of replacing conventional electronic-based AI engines, but aims rather to engender applications where photonics can offer certain benefits over their electronic counterparts. This mainly expands along inference applications, since inference comprises the most critical process in defining the power and computational resource requirements in certain applications such as modern Natural Language Models (NLP), where inference workloads are estimated to consume 25x-1386x higher power than training [43]. Other deployment scenarios, include latency-critical applications that are related to cyber-security in DCs [37], non-linearity compensation in fiber communication systems [39], acceleration of DNN's matrix multiplication operations at 10s of GHz frequency update rates [25], decentralization of AI input layer from core AI processing for edge applications [44] and finally to provide solutions to non-linear optimization problems in e.g. autonomous driving and robotics [40].

In this tutorial, we aim to provide a comprehensive understanding of the underlying mechanisms, technologies and training models of PNNs, highlighting their distinctive advantages and addressing the remaining challenges when compared to conventional electronic approaches. This tutorial forms the first attempt towards addressing the field of PNNs for DL applications within a hardware/software co-design and co-development framework: with the emphasis being on integrated PNN deployments, we define and describe the PNN fundamentals taking into account both the underlying chip-scale neuromorphic photonic hardware as well as the necessary optics-informed DL training models. The paper is structured as follows: In section 2 we introduce the basic definitions and requirements for NN hardware, analysing the basic NN building blocks (artificial neuron, NN models) as well as the main mathematical operations required for the hardware implementation of NNs i.e., multiply and accumulate (MAC) and matrix-vector-multiplication (MVM) operations. The same section provides also an intuitive analysis on bit resolution and energy efficiency trade-offs of analog photonic circuits, discussing the advantages and opportunities of PNNs. In sections 3 and, a review in the basic computational photonic hardware technologies is provided, presenting a summary of photonic MVM architectures and weight technologies in Section 3 and activations functions in Section 4. Finally, Section 5 is devoted to the challenges and requirements in the photonic DL training sector, providing a solid definition of optics-informed DL models and summarizing the relevant state-of-the-art techniques and demonstrations.

2. **Basic Definitions and Requirements for Neural Network Hardware**

Merging photonics with neuromorphic computing architectures requires a solid knowledge of the underlying NN architectures, building blocks and mechanisms. The most basic definitions and requirements are briefly described below:

*A. Artificial Neuron*

An artificial neuron comprises the main operation unit in a neural network, with the operation of the basic McCulloch-Pitts neuron model [45] being mathematically described by $y = \varphi(\sum W_i x_i + b)$, where $y$ is the neuron output, $\varphi$ is an activation (non-linear) function, $x_i$ is the $i^{th}$ element of the input vector $x$, $w_i$ is the weight factor for the input value $x_i$ and $b$ is a bias. The linear term $\sum W_i x_i$ represents the weighted addition and is typically carried out by the so-called linear neuron part, which comprises i) an array of axons, with every $i^{th}$ axon denoting the transmission line that provides a single $x_i \times w_i$ product, ii) an array of synaptic weights, with every $i^{th}$ weight $w_i$ located at the $i^{th}$ axon, and iii) a summation stage. The nonlinear neuron part comprises the activation function φ, with rectified linear unit (ReLU), sigmoid, pooling etc. being among the most widely employed activation functions in current DL applications [46].

For a layer of $M$ interconnected neurons, the output of these neurons can be expressed in vector form as: $y = (W \times x + b)$, where $x$ is an input vector with N elements, $W$ is the $N \times M$ weight matrix, $b$ is a bias vector with M elements, where $y$ is a vector made of $M$ outputs. Figure 1 (a) depicts a schematic layout of a biological neuron that can be mathematically described via an artificial neuron shown in Fig. 1(b), where the dendrites correspond to the weight signals, nucleus to the summation and activation function and axon terminals are responsible for providing the inputs to the next neuron, while Fig. 1 (c) depicts the resulting layout when utilizing artificial neuron to structure a DNN with a single input layer, a single output layer and one or more hidden layers.

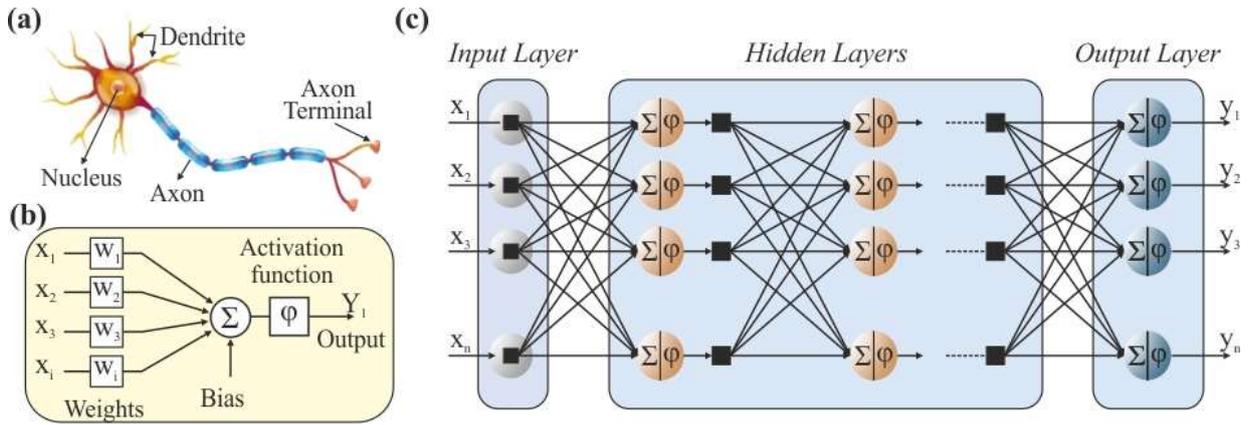

Fig. 1: (a) Schematic representation of a biological neuron (b) Basic model of artificial neuron comprising, a linear part (Σ) and a non-linear part (φ). Its fundamental operation are (i) the weighted addition (or synaptic operation) (ii) the non-linear activation function. (c) Example of a multi-layer NN comprising several neurons, one input layer, several hidden layers and one output layer.

*B. Neural Network models*

Part of the unprecedented success of NNs in tackling complex computational problems can be attributed to the plethora of NN models, capable of uniquely synergizing several hundred or up to billions of artificial neurons into versatile computational building blocks. In this section, we will give an overview of several NN models based both on their popularity and success in resolving standardized benchmarking problems, as well as their compatibility with hardware implementation in silicon photonic platforms.

NN models can be broadly classified in different categories based on their:

- <u>Data Flow pattern</u>. Considering the direction of the information flow, NN models can be grouped in two categories: In Feed-forward NNs the signals travel exclusively to one direction, usually from left to right, while in Feed-back NNs, the signals travel in both directions allowing neurons to receive data from neurons belonging to subsequent or even the same layer. Figure 2 (a)- (e) depicts five popular types of NN models, grouped based on their data flow in Feed-forward and Feed-back implementations, with the latter being mostly utilized for resolving temporal and ordinal workloads as the network effectively retains memory of the previous samples.

- Interconnectivity. The interconnection density between neurons of subsequent layers or even the same layer, can be used to classify NN models in dense and sparse implementations. Figure 2 (a) depicts a typical DNN model, where each neuron of the first layer is connected to all the neurons of the subsequently layer, usually denoted as a Fully-connected layout, while the neurons of the second layer are interconnected to only two neurons of the subsequent layer, corresponding to a sparse layout. While high-interconnectivity density allows the NN to extract more complex relationships between the input data, the cost associated with the increasing number of weights scaling with a $O(N^2)$ complexity for a $NxN$ interconnectivity, promotes the use of sparse models.

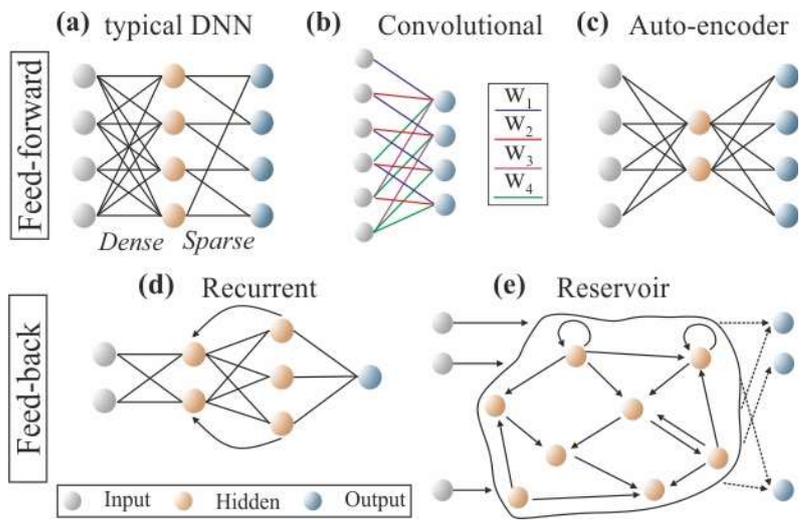

**Fig. 2:** Different types of Neural Network Models categorized in (a)-(c) Feed-forward and (d)-(e) Feed-back configurations

- Structural Layout. Employing a specific layout can enhance neural networks models with unique attributes. A typical example of such a model, specifically a single layer of a convolutional NN [47] is illustrated in Fig. 2 (b). This architectural approach, widely employed in image recognition task due to its spatio-local feature extraction capabilities, promotes weight re-use and as such relaxes the computational requirements, through applying the same weight kernel, i.e. a set of weight values, across the input data values. Another typical NN layout, depicted in Fig. 2 (c) is an NN autoencoder, a model associated with data encryption due to its data compression layout that effectively reduces the dimensionality of the input data in its central layers, boosts wide employment in non-linearity compensation in optical communications [42]. Finally, Fig. 2 (d) illustrates the most common Feed-back NN model called recurrent, while Fig. 2 (e) depicts a special type of recurrent NNs typically denoted as reservoir computing [48], where a fixed connectivity recurrent layer is placed between the input and the output layer. The relaxed training requirements, as only the output layer has to be trained along with the ease of constructing time-delayed reservoir circuitry in Silicon Photonics platforms, has led to impressive demonstrations in optical channel equalization applications.

### C. MAC and MVM operations

As highlighted in the previous subsections A and B, two types of mathematical operations are required for the hardware implementation of multi-layer NN models: (i) the linear MAC operation that constitutes the weighted summation at a single neuron ingress (ii) the non-linear activation function employed at the neuron's egress. Given that a N×N neural network layout comprises N neurons with N weighting elements per neuron, the two operations scale with $O(N^2)$ and $O(N)$ complexity, respectively. As such, MAC operations comprise the most significant computational burden and are usually correlated with the computational capacity of the NN model [49]. A single MAC operation calculates the product of two numbers and adds the result to an accumulator. Defining $a$ as the accumulation variable that holds the weighted input sum $\sum_{n=1}^{i-1} w_n \times x_n$, the operation can be described by the following form:

$$a_i \leftarrow a_{i-1} + (w_i \times x_i) \qquad (1)$$

A single artificial neuron with $N$ inputs, creates a weighted sum that can be broken down into a series of $N$ parallel MAC operations, while a fully connected neural layer with $M$ interconnected neurons and $N$ inputs per neuron supports a total number of $M \times N$ parallel MACs. A typical digital electronic MAC unit layout that realizes the mathematical formula of (1) is depicted in Fig. 3 (a), showing that the partial 2$N$-dimensional weighted sum, stored by the accumulator, is then fed back to the summation circuit for being added with the subsequent partial weighted sum produced by the next $N$ input and $N$ weight values. Figure 3 (b) illustrates an example of a MAC operational unit when implemented in the analog electronic domain, where an input vector x is imprinted in the electrical domain through the use of DACs and subsequently broadcasted to an array of [i, j] synaptic weights implemented through variable resistive elements arranged in a Xbar configuration [50], [51]. By controlling the impedance of the variable resistive elements, the electrical current emerging to every Xbar's column output, provides based on Kirchhoff's law the weighted summation of the column inputs [51], e.g.: $y_1 = \sum_{k=1}^{i} x_k * w_{k,1}$. Careful examination of the two MAC implementations can provide us with some significant insight into the differences between analog and digital computing:

- <u>Value Representation and information density</u>. Digital implementations use discrete values of physical variables, employing typically two discrete levels that are correlated with the upper and lower switching voltage of a transistor and are usually denoted as 0 and 1. On the other hand, analog computing employs values across the whole range of physical variables, allowing in this way for the representation of several equivalent bits of information at the same time unit. A direct consequence of this value representation form is the required noise robustness of the computational system that will be discussed in more detail in the following subsection, especially for optical implementations.

- <u>Computational primitives.</u> While digital computing is solely based on the mathematics and respective deployments of Boolean logic-based circuitry, analog computing can employ the physical laws of the underlying hardware, e.g. capacitors, resistors [52] to implement a variety of mathematical operations, unlocking a quiver of functionalities described by the exploited physical phenomena.

- <u>Latency.</u> Given the large number of devices required to implement a specific mathematical operation using Boolean logic, e.g., a digital computational building block implementing 8-bit parallel multiplication requires ~3000 transistors [53]. This forms a latency-critical computational path that is defined by the maximum register-to-register delay and effectively limits the maximum achievable operating frequency and, as such, the achieved latency [54]. This has led to the adoption of multi-threading and multi-core setups for parallel processing in modern computing systems, investing in architectural innovations towards system acceleration. On the other hand, analog systems are inherently built as parallel computational systems, giving them a significant edge in latency critical tasks, while requiring ~500x fewer components [53], on average, than digital electronic circuits for multiplication operations.

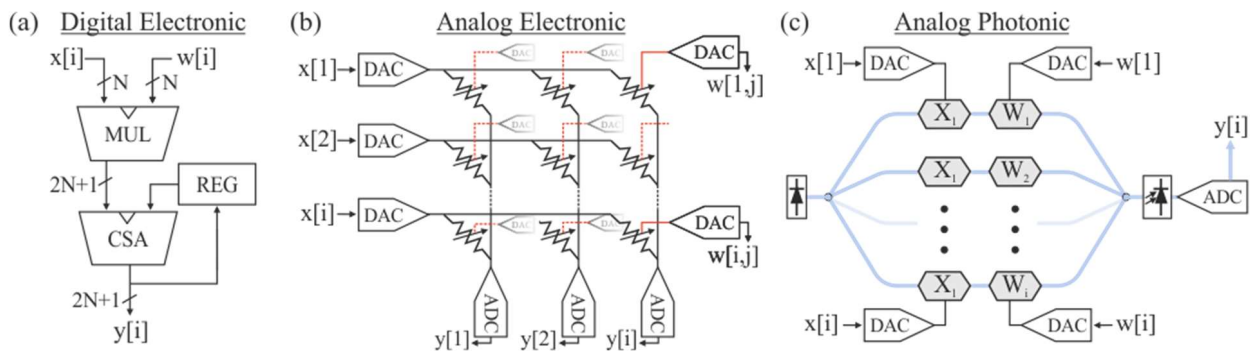

**Fig. 3:** Implementation of MAC operations in (a) Typical digital accelerators (b) Analog electronic approach employing Kirchhoff's law and a Xbar configuration (c) Indicative analog photonic approach employing coherent inference of light-beams.

These advantages, synergized with the primitives of photonic devices, have fueled the rise of optical MVM hardware, with an indicative example of an analog photonic dot product implementation given in Fig. 3 (c). In this approach, the input and/or weight information is encoded in one of the underlying physical variables of the photonic system i.e., the amplitude, phase, polarization or wavelength of a light beam, while the physical primitives of optical phenomena are utilized for the mathematical operations: in this particular example, loss experienced during the transmission of light via the weight-encoding physical system provides the multiplication operation, while interference of light waves is used for providing the summation mechanism. Harnessing the advantages of light-based systems i.e., multiple axes of freedom for encoding information in time, space and wavelength, low propagation loss, low electromagnetic interference and high-bandwidth operation hold the credentials to surpass analog electronic deployments in large scale photonic accelerators [55]. It is noteworthy, though, that both the electronic and photonic analog compute engines necessitate the use of Digital-to-Analog (DAC) and Analog-to-Digital (ADC) modules for interfacing NN input and output modules with the digital world.

### D. Precision

Migrating MAC operations from digital circuitry, where high-precision (i.e., 16-, 32- or 64-bit) floating point representations are utilized, to the analog domain, necessitates a basic understanding of the physical representation and energy-efficiency tradeoffs of analog photonic circuitry. Given the continuous nature of analog variables, as opposed to the usually two-level discretized variables in digital systems, representing high-precision numerical quantities in an analog system necessitates significantly higher signal-to-noise ratios (SNR). This requirement shapes an optimal bit resolution/energy efficiency operational regime for analog photonic computing systems [56]. In this subsection, we will discuss the precision limitations of analog photonic computing, outlining its optimized operational trade-offs in the shot-noise limited regime versus state-of-the-art digital MAC circuitry.

In digital implementations, where scaling the bit resolution is achieved by parallel circuitry operating at discrete binary levels, we can make the simplified assumption [53] that the power consumption of a digital MAC scales linearly with the bit-resolution such as:

$$P_D = b_{dig} \times P_{D-single-bit} \tag{2}$$

, where $P_{D-single-bit}$ is the power consumption of a single-bit MAC operation and $b_{dig}$ is the bit resolution. For photonic implementations, we use the correlation between the achieved bit resolution and the standard deviation of the system's total noise level as this has been defined in [57]:

$$b_a = log_2\left(\frac{I_{max} - I_{min}}{\sqrt{12} \times \sigma_{TOTAL}} + 1\right) \tag{3}$$

, where $I_{max} - I_{min}$ defines the range between maximum and minimum electrical current values generated at the photodiode output and $\sigma_{TOTAL}$ is the standard deviation of the total noise of the photonic link, under the generally valid assumptions that the link is dominated by additive white gaussian noise (AWGN). Assuming that the link operates at the shot noise limit of the photodiode, the total noise of the system equals the shot noise and we have:

$$\sigma_{TOTAL} = \sigma_{shot} = \sqrt{2 \times h \times v \times I_{avg} \times B} \tag{4}$$

,where $h$ is the Planck constant, $B$ the employed bandwidth, $v$ the Lightwave frequency, ($\lambda = 1550\ nm$ or $v = 193.41\ THz$) and $I_{avg}$ is the average electrical current generated at the photodiode output, which relates to the average optical power $P_{avg}$ that enters the photodiode via $I_{avg} = RP_{avg}$, with $R$ being the photodiode responsivity. It should be noted that the aforementioned calculation is an approximation of the actual shot noise of the photonic link, as its value is dependent on the number of photons reaching the receiver in any given computing interval [55], [56]. Taking into account that $I_{max} - I_{min} = R(P_{max} - P_{min}) = R \times OMA$, with $OMA$ denoting the Optical Modulation

Amplitude, and assuming that the input signal has a duty cycle of 50% and an infinite extinction ratio (ER), the $OMA$ turns to be equal $OMA = 2 \times P_{avg} = \frac{2 \times I_{avg}}{R}$, with $I_{max} - I_{min} = 2 \times I_{avg}$. Considering an ideal optical MAC unit in order to calculate the theoretical limit of optical energy efficiency, we assume only the use of unitary and lossless layouts, where a link, performs in principle, lossless MAC operation [58] and is powered by a laser that consumes an average power of $P_{laser}$ and has a wall-plug efficiency of $a = 0.2$, the average optical power emitted by the laser will be $a \times P_{laser}$ and will be required to be greater or equal to $P_{shot}$ when operation at the shot-noise limit is required. Finding $\sigma_{TOTAL} = \sigma_{shot}$ via equation (3) and then using the resulting expression to replace $\sigma_{shot}$ in eq. (4), while replacing $I_{avg}$ with $R \times P_{avg}$ and requesting $P_{avg} = P_{shot}$ due to the operation at the shot-noise limit, the $P_{shot}$ and, consequently, the consumed laser power $P_{laser}$ can be calculated as:

$$P_{laser} \geq \frac{P_{avg}}{a} = \frac{P_{shot}}{a} = 3.85 \, aJ \times (2^{b_a} - 1)^2 \times (1/R) \times B \qquad (5)$$

Considering then a 1:1 relationship between available bandwidth, expressed in GHz, and obtained MAC/s performance and dividing both parts of eq. (5) with $B$, we can transform equation (5) to the shot-noise limited energy efficiency per MAC described in equation (6):

$$E_{shot} \, (J/MAC) \geq 3.85 \, aJ \times (2^{b_a} - 1)^2 \times (1/R) \qquad (6)$$

Figure 4 (a) puts in juxtaposition the achieved energy efficiency of a digital MAC circuit, using equation (2) for a reference value of 46 fJ for 8-bit MAC [59], with the shot-noise limited energy efficiency of a single photonic circuit calculated by equation (6) when assuming a unity responsivity ($R = 1$). Another important metric that could be utilized for the aforementioned comparison is the Landauer Limit [60], that effectively defines the minimum energy required for a digital irreversible computation and as such could be employed as the theoretical minimum energy for digital computation, with an analysis and relevant metrics accessible in the supplementary material of [61]. While a more detailed discussion and related advantages of the achieved energy efficiency of a photonic accelerator will be provided in the next section, it becomes evident that harnessing the analog-architecture derived advantages of photonic implementations, the bit-resolution of the photonic accelerator will have to range in lower bitwidths than its digital equivalent. Moreover, as analog systems encode the data information along single physical variables, they have to migrate from floating-point to fixed-point representations. This is dictated by both the lack of bit-resolution depth that would allow splitting the mantissa and exponent part of the represented number, as well as the nature of computation that requires physical number representations. In order to ease the understanding of the two different representation-schemes, Fig. 4 (b) schematically describes the two approaches.

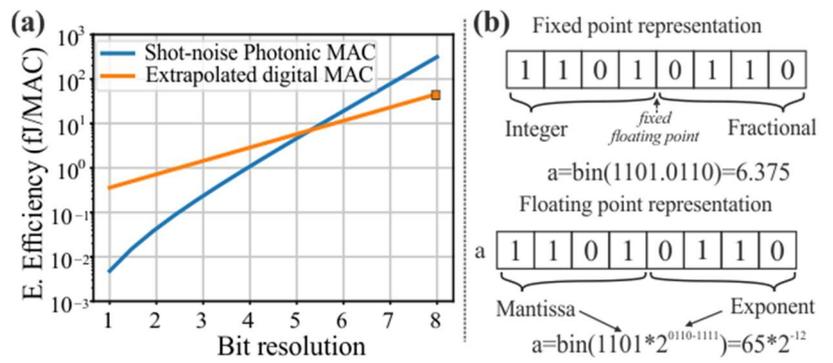

**Fig. 4:** (a) Comparison of Digital MAC energy efficiency versus shot-noise limited photonic MAC. Even though this graph provides insight into the optimized bitwidth of photonic implementations, it should be noted that photonic accelerators can surpass this efficiency due to favorable scaling laws as will be discussed in the next subsection. (b) Example of fixed vs floating point arithmetic.

Fortunately for analog MAC implementations we can also take advantage of the reduced, as compared to digital, precision required at the circuit's output. More specifically, considering an analog MAC implementation where #*N* *k*-bit signals are summated, the full digital resolution at the chip's output would be defined through:

$$b_{f-digital} = k + log_2 N \tag{7}$$

In a photonic computing scenario, assuming a layout similar to the one depicted in Fig. 3 (c) with a lossless weight matrix, the full digital precision can be only retained when a multiplication stage with a gain of *N* is employed for compensating for the *1/N* splitting ratio of the laser source at the circuit's ingress. However, in analog implementations we retain the same bit-precision across the different neural layers and as such the output of the summation should have the same bitwidth as the input neuron values. Consequently, the required SNR at the analog photonic output would be lower than the full digital equivalent one, implying that we can keep the minimum optical power difference between adjacent bits (MOPB) constant, even when reducing the laser optical power. Defining this digital-to-analog precision loss [55], [56] as $a_{prec}$, we can highlight two interesting operational regimes, schematically illustrated in Fig. 5 (a) and (b). Specifically in Fig. 5 (a), a light beam, originating from a laser source and consuming an optical power of $P$, is split in an $1{:}N$ splitter ($N = 4$) into 4 equivalent beams that get subsequently modulated in $X_1$- $X_4$ optical modulators. With the 4 inputs having a bit resolution of $X_{res} = 2$, their summation, using the equation (7), has a full digital precision of $Y_{res} = 2 + log_2 4 = 4$ and the system corresponds to $a_{prec} = 1$. On the other hand, in Fig. 5 (b), we set the output bit resolution to $Y_{res} = 2$, and hence, assuming only positive weights and a lossless weight matrix, we can maintain the same MOPB at the output summation even when reducing the injected optical laser power to $P' = P/N = P/4$, while the systems now corresponds to a $a_{prec} = 4 = N$. A more thorough analysis is given in the appendix. In this context, NNs are uniquely suited for analog computing, as empirical research has shown that they can operate effectively with both low precision and fixed-point representation with inference models working nearly just as well with 4-8 bits of precision in both activations and weights—sometimes even down to 1–2 bits [62]. On top of that, bit precision in analog compute engines can be improved by incorporating in the NN training the idiosyncrasies and noise sources of the underlying photonic hardware, investing in this way in the so-called hardware-aware training or Optics-informed DL models [31]. Employing this approach, researchers have already showcased robust networks that can secure almost the same accuracy with noise-free digital platforms [63], while a more detailed discussion is included in Section 5.

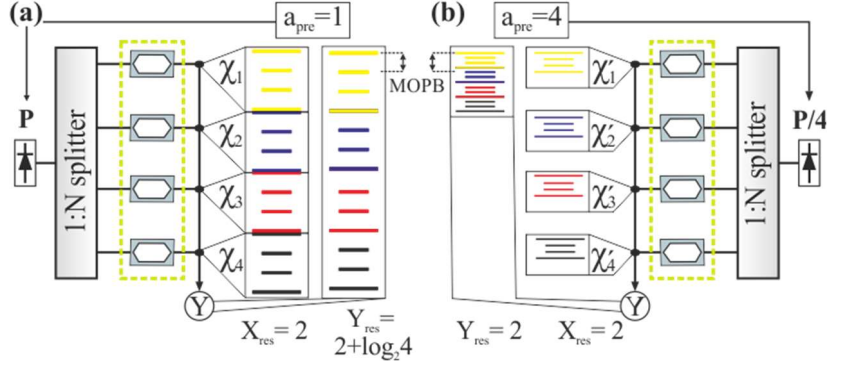

**Fig. 5:** Illustrative example of effect of output precision in photonic neural networks. # 4 2-bit inputs ($X_1$-$X_4$) are summated for two different digital-to-analog precision loss. (a) For full digital precision we have $\mathbf{Y_{res}} = \mathbf{4}$, (b) For same input and output precision bitwidths, $\mathbf{Y_{res}} = \mathbf{2}$. Assuming a lossless weight matrix and the same receiver sensitivity we can lower the input laser power by N ($\mathbf{P'} = \mathbf{P/4}$).

E. *Technology requirements for energy and area efficiency*

Energy efficiency is mainly dictated by the supported computational speed and the overall power consumed for the computations, which can be broken down into the power consumption of the laser source, fan-in, fan-out and weighting technology. On the same line, area efficiency gets determined by the computational speed divided by the overall footprint required for the computations, which depends, obviously, strongly on the size of the individual fan-in, weighting and fan-out structures. In an effort to determine the main parameters that affect the energy and area efficiency, we performed the following analysis for an $NxN$ neural layer, pictorially represented in Fig. 6. The $NxN$ neural layer comprises an $NxN$ weight matrix $W$, which gets multiplied by an $N{:}1$ input vector $X$ and yields an $N{:}1$ output vector $Y$. Assuming that every synaptic weight is implemented via a hardware module that consumes a power of $P_W$ watts and an area of $A_W$ mm², each input signal generation structure is realized by a hardware circuit that consumes a power of $P_X$ watts and an area of $A_X$ mm², the receiver circuitry that is employed for obtaining every output signal, consumes $P_Y$ watts and has a footprint of $A_Y$ mm² and the optical laser source consumes $P_{laser}$ watts, then the total power consumed equals:

$$P_T = NP_X + N^2 P_W + NP_Y + P_{laser} \tag{8}$$

Assuming an operation at $B$ MAC/sec compute rate per axon, then the total compute rate equals $N^2 B$ MAC/s, leading to an energy efficiency in J/MAC (or in MAC/s/watt) of:

$$\begin{aligned} E_{eff} &= \frac{NP_X + N^2 P_W + NP_Y + P_{laser}}{N^2 B} \\ &= \frac{P_X}{NB} + \frac{P_W}{B} + \frac{P_Y}{NB} + \frac{P_{laser}}{N^2 B} \end{aligned} \tag{9}$$

By carefully examining the contribution of the constituents of equation (9) we can conclude to the basic operational trade-offs of neuromorphic photonic accelerators. The first term encompasses the driving circuit of the input modulators, where its minimum required dynamic power consumption at a $B$ computational rate can be approximated

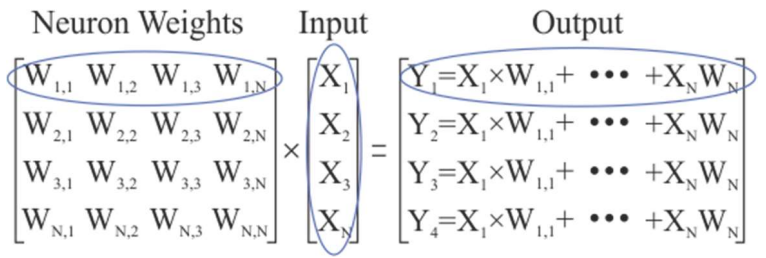

**Fig. 6:** Matrix-vector-multiplication of $NxN$ weight matrix and an $N{:}1$ input vector $X$, resulting to an $N{:}1$ output vector $Y$.

via their switching energy [64], defined in J/s as:

$$E_d = 0.25 \times C \times V^2 \times B \tag{10}$$

, where $C$, $V$ is the capacitance and the driving voltage of the input modulator, respectively. After merging equations (9), (10) and employing typical values ($C = 14\,fF, V_{pp} = 2\,V$) [65] for state-of-the-art electro-absorption modulators (EAMs) while excluding, at this point, their static energy consumption, it can be derived that:

$$\frac{P_x}{NB} = \frac{(E_d)}{NB} = \frac{(0.25 \times C \times V^2 \times B)}{NB} \approx \frac{14\,fJ}{N} \tag{11}$$

The second term $\frac{P_w}{B}$ corresponds to the driving circuit of the weight matrix, assuming stationary operations, where we can discern some typical operating regimes based on the deployed technology and compute rate, as given in Table 1 below:

**Table 1.** Typical weight technologies in silicon photonic accelerators

| Technology | Compute Rate B (MAC/s) | Static consumption (W) | Efficiency (J/MAC) |
|---|---|---|---|
| TO PS [66] | 10-50 E+09 | 12 E-03 | 2.4-12 pJ/MAC |
| Insulated TO PS [66] | 10-50 E+09 | 4 E-03 | 0.4-2 pJ/MAC |
| EAMs [65] | 10-50 E+09 | 2-20 E-06 | 0.4-20 fJ MAC |
| Non-volatile PCMs [26] | 10-50 E+09 | ≈0 | ≈0 pJ/MAC |

The EAM static consumption is calculated through:

$$E_{static} = (1/2) \cdot P_{in_{EAM},optical,mW} \times R \times V_{static} \tag{12}$$

considering an average static bias voltage of $V_{static} = -1.5\,V$, corresponding to the mean value of a uniform distribution that ranges in the EAMs operating regime, i.e., [0V-3V], a responsivity of $R = 0.8\,A/W$ and $P_{in}$ ranging from -15 dBm to -5 dBm. Regarding the thermo-optic (TO) phase shifter (PS) [66] we assume a uniform distribution of the weight values, corresponding to power distribution in the $P_0$-$P_\pi$ range with an average value of $P_{TO} = P_\pi/2$.

The third term $\frac{P_Y}{B}$ incorporates the power consumption of the receiver circuitry. Following a similar analysis to the transmitter part and for an output voltage of 0.5 V, compatible with current CMOS transistors, we can conclude to:

$$\frac{P_Y}{NB} = \frac{(E_d)}{NB} = \frac{(0.25 \times C \times V^2 \times B)}{NB} \approx \frac{1.25\ fJ}{N} \tag{13}$$

The final term $\frac{P_{laser}}{N^2 B}$ corresponds to the power consumption of the optical laser source. Similarly to the analysis provided in [55], the input optical power reaching the receiver circuitry should be:

(i) Higher than the accelerator's noise energy. In this context, following the analysis of the previous subsection for the shot-noise limited optical power and considering an $NxN$ neural layer, with (a) a power splitting ratio of $N^2$, implying that we have to multiply the output power by $N^2$ to compensate the input and column splitting stages, and (b) a digital precision loss of $a_{prec} = N$, the shot-noise limited optical power can be calculated using the following equation, which forms actually a more detailed representation of the laser power calculated in eq. (5), where however, the digital precision loss and the compensation loss factor are also taken into account:

$$P_{laser\_shot} = 3.85\ aJ \times (2^{b_a} - 1)^2 \times (1/R) \times B \times \frac{1}{a_{prec}} \times N^2 \tag{14}$$

, which makes the constituent term of equation (9) to equal

$$\frac{P_{laser\_shot}}{N^2 B} = \frac{3.85\ aJ \times (2^{b_a} - 1)^2 \times (1/R) \times B \times \frac{1}{N} \times N^2}{N^2 \times B}$$
$$= 3.85\ aJ \times (2^{b_a} - 1)^2 \times \frac{1}{N} \tag{15}$$

, assuming a responsivity $R = 1$ A/W.

(ii) Sufficient to generate the minimum required electrical charge at the receiver that can drive the subsequent node of the next NN layer [67]. With the photonic accelerator operating at 1550 nm and assuming a photodetector with $C_d = 1\ fF$ [68], a $C_i = 200\ aF$, 1um wire with an interconnect capacitance of 200 aF/um [67] and a required output voltage of $V_{out} = 0.5\ V$ [69], the minimum optical power required can be calculated, following the same convention of $N^2$ splitting loss and $N$ digital precision loss:

$$V_{out} = \frac{P_{laser\_switch}}{a} \times \frac{e}{h \times v} \times (C_d + C_i) \times \frac{1}{a_{prec}} \times \frac{1}{N^2} \times B$$

$$P_{laser\_switch} = V_{out} \times (C_d + C_i) \times \frac{h \times v}{e} \times \frac{1}{N} \times N^2 \times B \tag{16}$$

, that concludes for the fourth term of the energy efficiency to

$$\frac{P_{laser\_switch}}{N^2 B} = \frac{V_{out} \times (C_d + C_i) \times \frac{h \times v}{e} \times \frac{1}{N} \times N^2 \times B}{N^2 \times B} = 2.4\ fJ \times \frac{1}{N} \tag{17}$$

Here, it should be pointed out, that this interconnect capacitance $C_i$ suggests monolithic integration approach or a very intimate proximity of the photonic chiplet to the respective electronic chiplet. More traditional integration approaches will enforce higher interconnect capacitances and significantly increase the required energy, with an interesting analysis provided in [70] Combining all the terms in a single efficiency equation we can conclude to :

$$E_{eff} = \frac{P_x}{NB} + \frac{P_w}{B} + \frac{P_Y}{B} + \frac{max\ (P_{laser_{shot}}, P_{laser_{switch}})}{N^2 B}$$

$$E_{eff} = \frac{5\ fJ}{N} + \frac{P_w}{NB} + \frac{1.25\ fJ}{N} + \frac{1}{N} \times max\ [3.85\ aJ \times (2^{b_a} - 1)^2,\ 2.4\ fJ] \tag{18}$$

This highlights that energy efficiency improves with:

- Increasing $N$, implying that the energy consumed for generating and receiving the input and output signal, respectively, is optimally utilized when the same input and output signals are shared along multiple matrix multiplication, or equivalently, neural operations. With current neuromorphic architectures being radix-limited by maximum emitted laser power [71], loss-optimized architectures are required for allowing high circuit scalability and harnessing the advantages of photonic implementations.

- Increasing B, that has a predominant effect in reducing energy consumption especially when using high-power consumption weight nodes i.e., currently widely employed thermo-optic heaters dominate the energy efficiency reaching up to ~1 pJ/MAC.

- Operating in an optimized bit resolution energy regime as highlighted in the 4[th] constituent of equation (18). As we can observe, the order of magnitude difference between the shot-noise limited and minimum switching energy contributions, has a threshold point at around 4.5 bits, implying that a careful examination of the underlying technology blocks and an optimized operational regime can significantly improve the energy consumption.

Following a similar analysis for the computational area efficiency, the total area consumed by this circuit equals $NA_X + NA_Y + N^2 A_W$ mm², suggesting an area efficiency in MAC/s/mm² of

$$F_{eff}(MAC/s/mm^2) = \frac{N^2 B}{NA_X + NA_Y + N^2 A_W} = B\left(\frac{N}{A_X} + \frac{N}{A_Y} + \frac{1}{A_W}\right) \quad (19)$$

revealing a positive linear relation with $B$ and relative gains for high accelerator radices.

## 3. Integrated Photonic Matrix-Vector-Multiply Architectures

### A. Coherent MVM architectures

In this section we initially investigate the architectural categories of integrated PNNs and then we delve deeper into their individual building blocks providing the recent developments on the photonic weight technologies as well as the non-linear activation functions implementations. Depending on the mechanism of information encoding and the calculation of linear operations, integrated PNNs can be classified into three broad categories: coherent, incoherent and spatial architectures.

Coherent architectures harness the effect of constructive and destructive interference for linear combination of the inputs in the domain of electrical field amplitudes, requiring just a single wavelength for calculating the neural network linear operations. The principle of operation of coherent architectures is pictorially represented in Fig 7 (a), while Fig. 7 (b)-(d) illustrate indicative coherent layouts that have been proposed in the literature and will be comprehensively analysed in this tutorial. The first linear neuron realized in this manner has been proposed in [29], with its core relying on optical interference unit realized through cascaded MZIs in a singular value decomposition

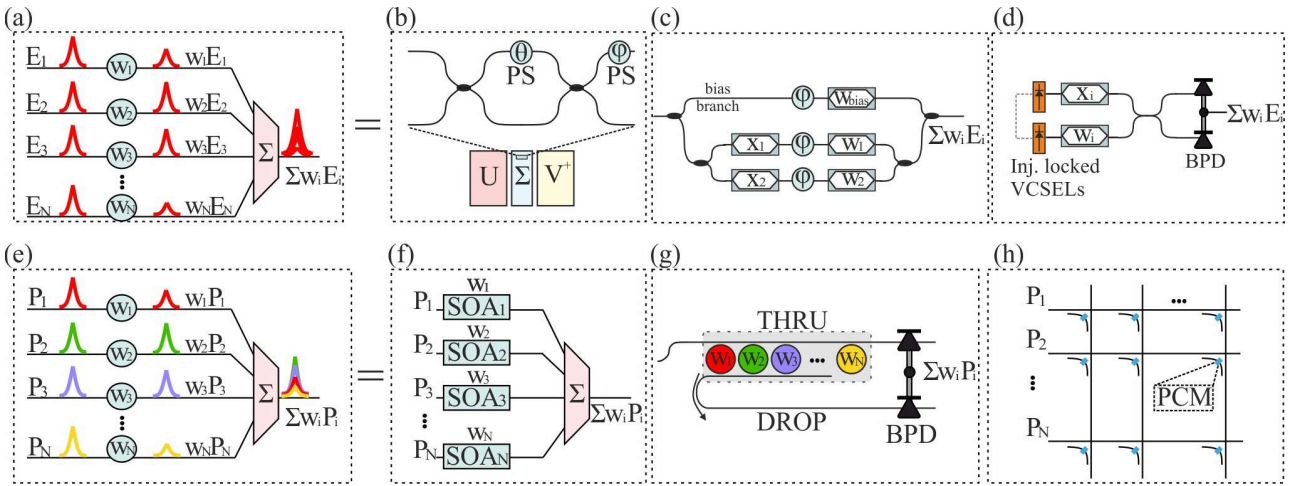

**Fig. 7**: (a) Coherent PNN linear architectures that are based on electric field summation, (b)-(d) indicative layouts from the literature that follow the coherent principles, (e) Incoherent PNN linear architectures that are based on WDM optical power addition, (f)-(h) indicative layouts from the literature that adopt the incoherent concept.

(SVD) arrangement [72], as per Fig. 8 (a). The SVD approach assumes decomposition of arbitrary weight matrix $W$ to $W = USV^\dagger$, where $U$ and $V$ denoting unitary matrices, $V^\dagger$ being the conjugate transpose of $V$, and $S$ being a diagonal matrix that carries the singular values of $W$. Therefore, this scheme rests upon the factorization of unitary matrices that in the photonic domain have mainly based on U(2) factorization techniques employing $2 \times 2$ MZIs [73].

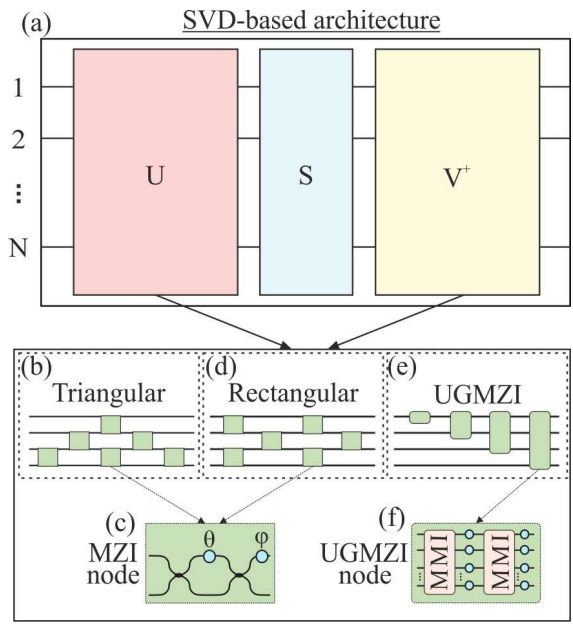

**Fig. 8**: (a) Singular Value decomposition scheme, (b) Triangular mesh implemented by Reck et.al, (c) Conventional Mach-Zehnder Interferometer with an additional PS element at one of the two outputs (d) Rectangular mesh proposed by Clements et. al, (e) UGMZI architecture proposed by Tsakyridis et. al., (f) Individual UGMZI node that operates as an $NxN$ beam splitter.

In this regime, back in 1994, Reck et. al., [74] proposed the first optical unitary matrix decomposition scheme, the so-called triangular mesh shown in Fig. 8 (b), using $2 \times 2$ MZIs as the elementary building block, illustrated in Fig. 8 (c). Recently, this layout has been optimized by Clements et. al., [75], introducing the rectangular mesh of $2 \times 2$ MZIs, depicted in Fig. 8 (d), that is more loss-balanced and error-tolerant design than Reck's architecture. Both layouts necessitate $N(N-1)/2$ variable beam splitters for implementing any $N \times N$ unitary matrix, requiring, also, the same number of programming steps for realizing the decomposition. Although these U(2)-based architectures rely on simple library of photonic components that facilitate their fabrication, they suffer from several drawbacks, with the most important being the fidelity degradation. Fidelity corresponds to the measurement of closeness between the experimentally obtained and the theoretically targeted matrix values, denoting a quantity that declares the accuracy in implementing a targeted matrix in the experimental domain. Fidelity degradation in the U(2)-based layouts originates from the differential path losses imposed by the non-ideal lossy optical components [76].

On top of that, U(2)-based layouts cannot support any fidelity restoration mechanism without altering their architectural structure or sacrificing their universality. Transferring these layouts in an SVD scheme towards implementing arbitrary matrices, the above effects exacerbate, as two concatenated unitary matrix layouts are required. In an attempt to counteract these issues, the authors in [77] proposed the universal generalized Mach-Zehnder interferometer (UGMZI)-based unitary architecture illustrated in Fig. 8 (e) and introduced a novel U(N) unitary decomposition technique [78] in the optical domain, that migrates from the conventional U(2) factorization by employing $N \times N$ Generalized MZIs (GMZIs) as the elementary building block. GMZIs serve as $N \times N$ beam splitters [79], [80], followed by $N$ PSs with each $N \times N$ beam splitter comprising two $N \times N$ MMI couplers interconnected by $N$ PS, as depicted in Fig. 8 (f). This scheme eliminates the differential path losses and hence, it can yield 100% fidelity performance by applying a simple fidelity restoration mechanism, which incorporates $N$ variable optical attenuators at the inputs of the UGMZI. Yet, this architecture heavily relies on MMI couplers with high number of ports in order to perform transformations on large unitary matrices, which are still a rather immature integrated circuit technology that is under development in current research fabrication attempts. Finally, the authors in [81] proposed the slimmed SVD-based PNN, where they have traded the universality for area and loss efficiency by eliminating one of the two unitary matrices, implying that they can implement only specific weight matrices.

Apart from SVD-based approaches, direct-element mapping architectures comprise also coherent layouts that employ a single wavelength and interference for calculating the linear operations. The mapping of the weight values to the underlying photonic fabric is bijective, meaning that each photonic node imprints a dedicated value of the targeted weight matrix without necessitating decomposition, minimizing this way the programming complexity. Fig. 9 illustrates the first coherent direct-element mapping architecture [76], implemented in a crossbar (Xbar) layout. In order to support both positive and negative weight values, this architecture requires the use of two devices per weight– an attenuator for imprinting the weight magnitude, proportional to $|W_i|$, and a PS for controlling the phase, i.e., the sign of the weight, $sign(W_i)$, enforcing 0 phase shift in case of positive, and $\pi$ phase shift in case of negative weights, resulting in $sign(W_i)|W_i| \times X_i$. The weighted inputs are linearly combined in $N:1$ combiner stage, constituted from cascaded Y-junction combiners, yielding the output electrical field proportional to $\sum_{i=1}^{N} X_i W_i$ which conceals the sign information in its phase. If compatibility with electrical nonlinearities is needed, the sign information of the signal emerging from the Xbar output can be translated from its phase to its magnitude by introducing an optional bias branch, which sets a constant reference power level that allows for mapping the positive/negative output field above/below the bias, as experimentally demonstrated in [82]. Xbar architecture, thanks to its loss-balanced configuration, can yield 100% fidelity performance, while its non-cascaded and one-to-one mapping connectivity, significantly improves the phase-induced fidelity performance, since the error is restricted only to a single matrix element. These benefits were experimentally verified in [83], [84] employing a 4x4 silicon photonic Xbar with SiGe EAMs as computing cells, while the NN classification credentials of this architecture were experimentally validated in [24], [25], using a 2:1 single-column Xbar layout that is capable to calculate the linear operations of the MNIST dataset at up to 50 GHz clock frequency with classification accuracy of >95%. In an effort to exploit the full potential of photonic platform, Xbar architecture can be equipped with WDM technology to further boost the throughput as has been proposed in [85],[86], realizing multiple output vectors at a single timeslot. Although the Xbar layout seems currently to be the optimal architectural candidate for PNNs, it requires careful and precise effort during circuit design in order to synchronize the optical signals that travel through different paths and coherently recombine at the output. Hence, optimum performance of the Xbar necessitates the employment of equal length optical paths whenever coherent recombination is required, suggesting that path-length difference has to be compensated during the photonic chip layouting.

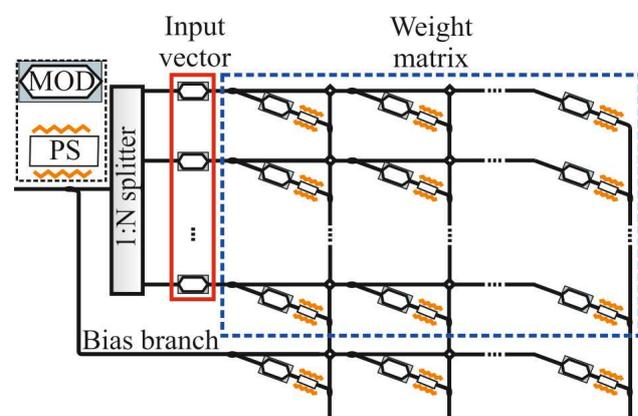

**Fig. 9**: Xbar architecture as a direct-elements mapping coherent layout.

Finally, a recent coherent demonstration in [87] exploits vertical-cavity surface-emitting lasers (VCSELs) for encoding, in *i*-time steps, both input vector and weight matrix, as shown in Fig 7 (d). Using injection locking mechanism between the deployed VCSELs, the phase coherency is retained over the entire circuit, allowing for the realization of the coherent amplitude addition at interference stage of each time-step. Matrix-vector products are realized by the photoelectric multiplication process in homodyne detectors, while a switched integrator charged amplifier is employed for the accumulation of the individual *i* products. Despite its simplicity, this architecture requires precise phase control over the individual VCSELs towards retaining phase coherency over the entire circuit, raising stability and scalability issues.

## B. Incoherent MVM architectures

Demarcating from coherent architectures, incoherent PNNs encode the NN parameters into different wavelengths and calculate the network linear operations by employing WDM technology principles and power addition. A pictorial representation of how incoherent architectures operate is given in Fig 7 (e), while some incoherent layouts that have been suggested in the literature and will be thoroughly examined in this tutorial are illustrated in Fig. 7 (f)-(h). The first implementation that follows this approach has been proposed in [88], when a team from Princeton initially demonstrated the so-called Broadcast-and-Weight architecture and then elaborated in more detail in [89]. Each input $x_i$ is imprinted at a designated wavelength $\lambda_i$, essentially making each channel $\lambda_i$ a virtual axon of a linear neuron, while all $N$ inputs ($\lambda_s$) are typically multiplexed together into a single waveguide when arriving to the linear neuron, as shown in Fig. 10. The main building block of this architecture is the micro-ring resonator (MRR) bank, consisting of $N$ MRRs that are embraced by two parallel waveguides and are responsible for enforcing channel-selective weighting values. Each MRR filter is designed such that its transfer function can be continuously tuned, ideally between the values of 0 and 1, achieving controlled attenuation of the signal's power at the corresponding $\lambda_i$ . The sign is encoded by exploiting path-diversity and balanced photodetection (BPD); assuming that an $a_i$ fraction of a signal at a certain wavelength exits via the THRU port of the respective MRR module and the remaining (1-$a_i$) part gets forwarded to the DROP port, the subtraction of the respective photocurrents at the BPD yields the weighting value $w_i = 2a_i - 1$ for this specific signal, which can range between -1 and 1 given that $a_i$ ranges between 0 and 1. With all different wavelengths leaving through the same DROP and the same THRU port and entering the same BPD unit, the BPD output provides the total weighted sum of WDM inputs. This architecture allows for one-to-one mapping of the weighting values into the MRR weight bank alleviating the programming complexity, yet it comprises a rather challenging solution since it necessitates the simultaneous operation and precise control of various resonant devices, raising issues in its scalability credentials. An alternative incoherent architecture is proposed by the authors in [26], demonstrating a PNN that follows the photonic in-memory computing paradigm where the weighting cells are realized through PCM-based memories. This approach exploits the non-volatile characteristics of the PCM devices consuming, in principle, zero power consumption when inference operation is targeted, meaning that the weights of the NN do not have to be updated and thus are statically imprinted in the PCM weighting modules. This architecture utilizes an integrated frequency comb laser to imprint the multiple inputs of the NN, with each comb line corresponding to a dedicated NN input value. The multi-wavelength signals after the PCM-based weighting stage that follows the layout depicted in Fig. 7 (h), are incoherently combined to a photodiode (PD) in order to produce the linear summation. Although this architecture minimizes the memory movement bottleneck, it requires i) precise design to timely synchronize the multi-wavelength signals at each PD and ii) broad wavelength spectrum of frequency comb laser for implementing large scale NNs. An additional incoherent architecture is proposed in [28] and illustrated in Fig 11. The authors employ WDM input signals for imprinting the NN input vector, while the realization of the weight matrices is implemented via multiple semiconductor optical amplifiers (SOAs). They adopt the cross-connect switch principles used in optical communications for constructing the PNN and arrayed waveguide gratings (AWGs) for multiplexing/demultiplexing the signals as well as for reducing the out-of-band accumulated noise of the SOAs. Although it comprises a promising solution towards implementing large scale PNNs, the deployment of multiple SOAs, as single stage weighting elements, trades the scalability credentials for increased power consumption. Finally, an alternative to the coherent/incoherent architectures has been proposed in [32], where the authors encode the N pixels of the classification image (NN input values) directly to the grating couplers through optical collimators, while the weight information of each NN input is imprinted through a dedicated PIN-based optical attenuator. Each weighted input is launched to a PD and the resulted photocurrents are combined to generate the linear weighted sum of the neurons. As opposed to the coherent and incoherent layouts, there is no requirement for the encoded signals to be in phase or in different wavelength, respectively, since every NN input is imprinted at a designated photonic waveguide/axon. This, however, necessitates multiple waveguides/axons for implementing a high-dimensional NN, imposing scalability limitations.

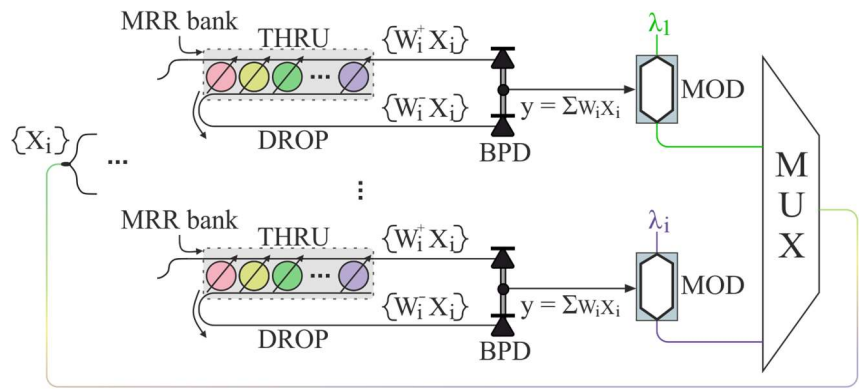

**Fig. 10**: Broadcast and weight architecture using microring resonator bank for realizing the weighting function.

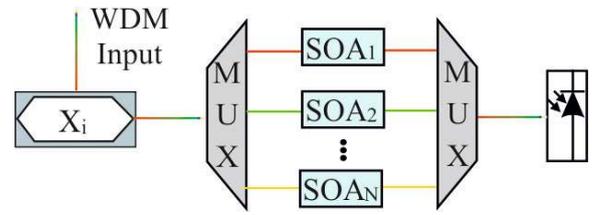

Fig. 11: Semiconductor optical amplifier-based incoherent photonic neural network.

From all previous implementations, it becomes easily evident that the main challenges and limitations of integrated PNNs relate to their scalability and thence to the hardware encoding of the vast amount of NN parameters into a photonic chip. In this direction, the authors in [90], [38], [91] introduced the optical tiled matrix multiplication (TMM) technique, shown in Fig 12, that follows the principles of general matrix multiply (GeMM) method adopted by modern digital AI engines [92]-[93] and attempting to virtually increase the size of the PNN without fabricating large photonic circuits. The rationale behind this concept is the following: the weight matrix and the input vector of an NN is divided into smaller tiles, whose dimension is dictated by the available hardware neurons. The rest tiles are unrolled in the time domain via time division multiplexing (TDM) and then are sequentially imprinted into the photonic hardware, allowing in this way for the calculation of matrix-multiplication operations of a NN layer whose dimension is higher than the one implemented on hardware. The resulting time-unfolded products, produced by the multiple tiles, need to be added together in order to form the final summation. For this reason, the authors in [91], [94], [95] utilized a charge accumulation technique either electro-optically using a low-bandwidth photodetector or electrically via a low-pass RC filter. Besides accumulation, this implementation allows for power efficient and low-cost ADCs, since it relaxes their sampling rate and bandwidth requirements. However, the employment of optical TMM and charge accumulation techniques in a PNN engender specific requirements that need to be addressed: i) both input vector-imprinting and weight-encoding modulators have to operate at the same data rate, ii) the number of time-unfolded products that will be accumulated is dictated by the deployed capacitance of the RC filter or the bandwidth of photodetector, implying that after a certain period, a capacitor voltage/photodetector power should be reset in order to store (e.g. to a local memory) the $1^{st}$ set of accumulated summation. The same process is repeated until the calculation of the total linear operations of PNN.

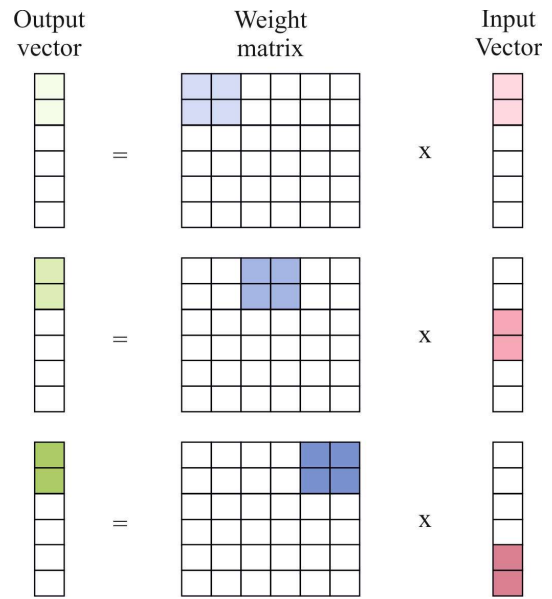

Fig. 12: Tiled matrix multiplication technique that firstly proposed as general matrix multiply method by modern digital AI engines.

4. **Neuromorphic photonic hardware technology**

   *A. Photonic weighting technologies*

Delving deeper into the individual PNN building blocks, we provide an overview of photonic technologies that can be promising candidates towards the realization of the NN weight imprinting into an integrated platform. As discussed previously, most PNN demonstrations focused on the weight matrix implementation rather than the NN input vector, since the number of weight values comprises the greatest contributing factor to the hardware encoding of the entire NN parameters. For example, assuming a fully-connected NN with topology of 10:10:5, the number of input values is 10 while the total weight values is 150, and this difference becomes more pronounced as the NN dimensions/layers increase. Hence, the selection of the photonic weight technology becomes crucial, as it implicitly indicates the size and energy efficiency of the PNN. The photonic weight technologies can be divided into two categories, depending on their volatile characteristics. Non-volatile devices can be used as memories by storing the NN weight values in a PNN and this information can be retained by statically applying ultra-low or even zero electrical power. These devices can either use memristors heterogeneously integrated with photonic microring resonators [96] or exploit physical phenomena such as phase change [26] and ferroelectricity [97] in order to store and retain the weight values. The employment of non-volatile memory elements is more suitable for equipping PNN inference engines, offering low-power weight encoding with high-precision, but, in turn, they impose challenges that are related to reconfiguration time, fabrication maturity, compactness and scalability. For example, PCMs that are mostly based on GST-based compounds, exhibit up to 5-bit resolution [98], but in turn, their reconfiguration time is restricted to sub-MHz regime, while in most demonstrations operate via optical absorption, limiting their deployment in large scale circuits. Ferroelectric materials such as Barium Titanate (BTO) have already validated its non-volatile credentials retaining its states over 10 hours [97]. However, to incorporate this device into a PNN, one aspect that still needs to be addressed and optimized is the footprint, since the required PS length for achieving pi phase shift is at least 1 mm [97], rendering the implementation of a large scale PNN rather challenging. On the other hand, when training applications are targeted or the TMM technique has to be applied for executing a high dimension neural layer over a limited PNN hardware, volatile devices take the lead over non-volatile materials since they offer dynamic weight update. Various TO MZI or MRR [27], [29], [30], [31], [99]-[101] devices have been proposed for weight data encoding due to their well-established and mature fabrication process as well as their high bit precision (up to 9-bit [101]), yet their reconfiguration time is limited to ms values.

Electro-optic devices such as micro-electro-mechanical systems (MEMS) [102], [103], EAMs [36], semiconductor optical amplifiers (SOAs) [28], ITO-based modulators [104]. Graphene-based phase shifters [105], Silicon p-i-n diode Mach-Zehnder Modulator (MZMs) [106] have already been demonstrated and potentially perform weighting functions exhibiting reconfiguration times in GHz regime trading however their performance in bit precision [107]. Therefore, the selection of the photonic weight technology heavily depends on the targeted NN application (inference, training) and its bit resolution requirements. Figure 13 puts in juxtaposition the power consumption and footprint of different photonic technology candidates for the realization of the weighting function for PNN implementations, highlighting also their speed capabilities/reconfiguration time.

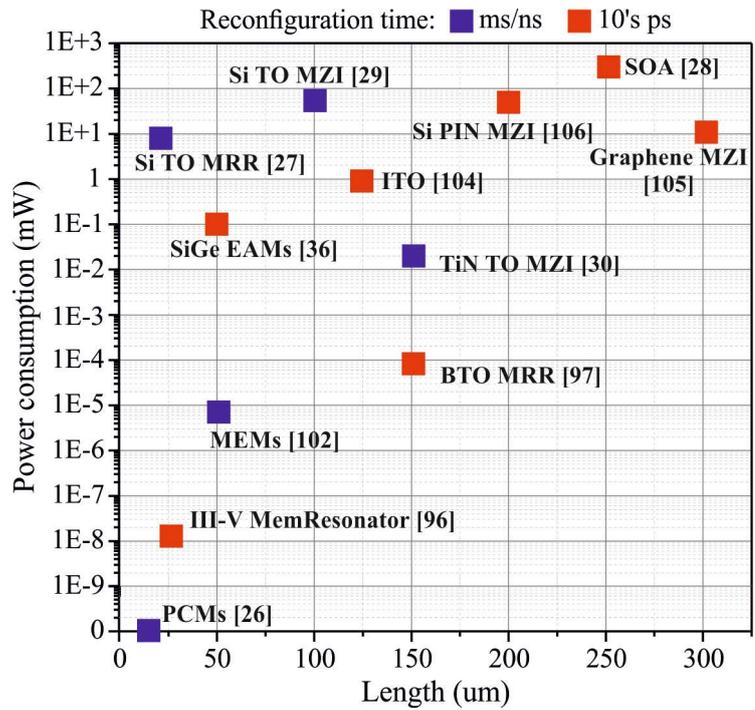

**Fig. 13**: Power consumption vs device length of various optical technologies for realizing weighting circuits. They are group in two different categories, depending on their reconfiguration time: blue square dots for low speed (ms/ns), red square dots for high-speed (10's of ps).

### B. Photonic activation functions

An indispensable part in the realization of an NN is the activation function, i.e., a non-linear function that is applied at the egress of the linear weighted summation. The non-linearity of the activation function allows the network to generalize better, converge faster, approximate non-linear relationships, as well as avoid local minima. Despite the relative relaxed requirements in the properties of the activation functions i.e., a certain degree of non-linearity and differentiability across the employed range [108], DNN implementations have been dominated, due to their higher performance credentials, by the use of the ReLU [109], PReLU [110] and variations of the sigmoid transfer function, including tanh and the logistic sigmoid [111]. This dominance has shaped photonic NN activation function circuitry objectives, targeting to converge to these specific electrical baseline functions performance at highest possible bit rate as well as achieve a certain level of SNR at their output to safeguard the scalability of the neural circuitry. Previous implementations of non-linearity in photonic NNs have been streamlined across three basic axes: (i) The simplest approach relies on applying the non-linear activation in the electronic domain. This was achieved through offline implementation in a CPU, following the opto-electrical conversion of the vector-matrix-multiplication product [29], by chaining an ADC to a digital multiplier and finally to a DAC [112], or by introducing non-linearity in the neuron's egress through a specially designed ADC [94]. Despite the simplicity and effectiveness of digitally applying the non-linear activation function, the related unavoidable digital conversion, induces, in the best case, a latency of several clock-cycles for every layer of the NN that employs one [29], [112]. Transferring this induced latency to a photonic NN accelerator would significantly decrease the achieved computation capabilities and as such its total performance credentials. (ii) The hybrid electrical-optical approach that relies on a cascade of active photonic and/or electronic components i.e., Photodiode—Amplifier-Modulator-Laser, with non-linear behaviour provided by the opto-electrical synergies, such as transimpedance amplifier (TIA), or by the non-linear behaviour of the photonic components (e.g. modulators) [113]-[118]. The hybrid electrical-optical approaches provide a viable alternative to digitally applied activation functions, but, in turn, the induced noise and latency originating from the cascaded optical to electrical to optical conversions still may impose a non-negligible overhead to the performance of the photonic NN. (iii) The all-optical approach based on engineering the non-linearities of optical components to conclude to practical photonic activation functions. In this context, different mechanisms and materials have been investigated, including among others gain saturation in SOAs [119], [120] absorption saturation [121], [122], reverse absorption saturation in films of buckyballs (C60) [122], PCM non-linear characteristics [123], [124], SiGe hybrid structure in a microring resonator [125], [126] and poled thin- lithium niobate (PPLN) nanophotonic waveguides [127]. All optical approaches seem to hit the sweet spot, between applicability and function, allowing time-of-flight computation and negating the need of costly conversions.

Finally, a recent trend and probably the most promising for realizing a complete PNN, comprises the development of programmability feature in both hybrid and all-optical approaches, where a single building block can realize multiple activation functions by modifying its operational conditions [115], [117], [118], [122], [124]. These implementations have mainly relied on the different non-linear transfer functions obtained by the same component, when altering its operational conditions through specific settings, e.g. DC bias voltage for a modulator, DC current for an SOA, gain of a TIA, input optical power and pulse duration for PCM etc. Therefore, by enabling reconfigurability in PNNs can pave the way towards implementing different AI applications/tasks without requiring any modifications in the underlying hardware.

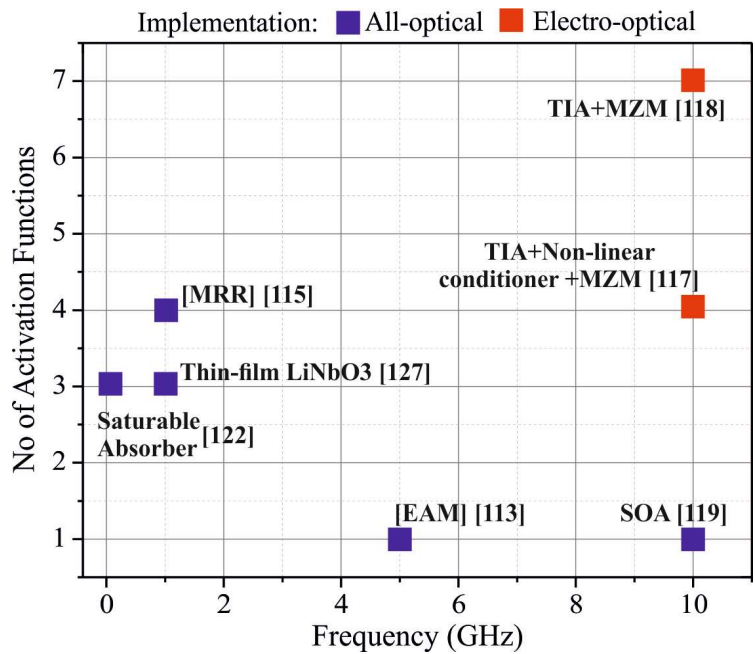

**Fig. 14**: Number of activation functions produced by a single device vs operating frequency. They are classified based on their implementation: Blue square dots for all-optical, Red square dots for electro-optical.

Yet, the programmability properties of the non-linear activation functions need to be combined with high-speed performance to comply with the frequency update rate of the execution of linear part. Figure 14 provides an overview of the devices that have been proposed for the implementation of NN activation functions, classifying them according to their implementation (all-optical, electro-optical), their speed performance and the number of activation functions that they can realize, while Table 2 summarizes the power consumption and area metrics of state-of-the-art activation function demonstrations.

**Table 2:** A summary of the power consumption and area metrics for the most advanced activation function demonstrations

| References | Power Consumption (mW) | Area (mm$^2$) |
|---|---|---|
| TIA+MZM [118] | 425 | 7.13 |
| TIA+Non-linear cond+MZM [117] | 400 | 0.625 |
| SOA [119] | 1640 | 9.1 |
| EAM [113] | 17 | N.A. |
| Thin Film LiNbO$_3$ [127] | 135×10$^{-3}$ | N.A. |
| Saturable Absorber [122] | 40×10$^3$ | 11.76×10$^3$ |
| MRR [115] | 0.1 | 25 |

5. **Optics-Informed Deep Learning Models**

Despite the significant energy and footprint advantages of analog photonic neuromorphic circuitry, its use for DL applications necessitates a unique software-hardware NN co-design and co-development approach for accounting for various factors that are absent in digital hardware and as such ignored in current digital electronic DL models [114]. These include among others: fabrication variations, optical bandwidth, optical noise, optical crosstalk, limited ER and non-linear activation functions that deviate from the typical activation functions used in conventional DL models, with all of them acting effectively as performance degradation factors [128]. In this context, significant research effort has been invested into incorporating the photonic-hardware idiosyncrasy in NN training models [129], engendering also a new photonic hardware-aware DL-framework. This reality has shaped a new framework for PNNs that should be eventually better defined as the NN field that combines neuromorphic photonic hardware with Optics-informed DL training models, using light for carrying out all constituent computations but at the same time using DL training models that are optimally adapted to the properties of light and the characteristics of the photonic hardware technology. The research field of hardware-aware DL training models designed and deployed for neuromorphic photonic hardware has led to the introduction of Optics-informed DL models [31], [37], [130], [131], a term that has been recently coined in [42] revealing a strong potential in matching and even outperforming digital NN layouts in certain applications [42].

Optics-informed DL models have to embed all relevant noise and physical quantities that stem from the analog nature of light and the optical properties of the computational elements into the training processing. In order to ease the understanding of the noise sources and physical quantities that impact a photonic accelerator and related NN implementation challenges, Figure 15 (a) exemplary illustrates the implementation of a single neuron axon over photonic hardware, along with the dominant signal quality degradation mechanisms. The input neuron data $x_i$ is quantized prior being injected in a DAC, whose bit resolution for the tens of GSa/s sampling rates required for photonic neuromorphic computing ranges around 4 to 8 bits [132], i.e. being significantly lower than the 32-bit floating point numbers utilized in digital counterparts.

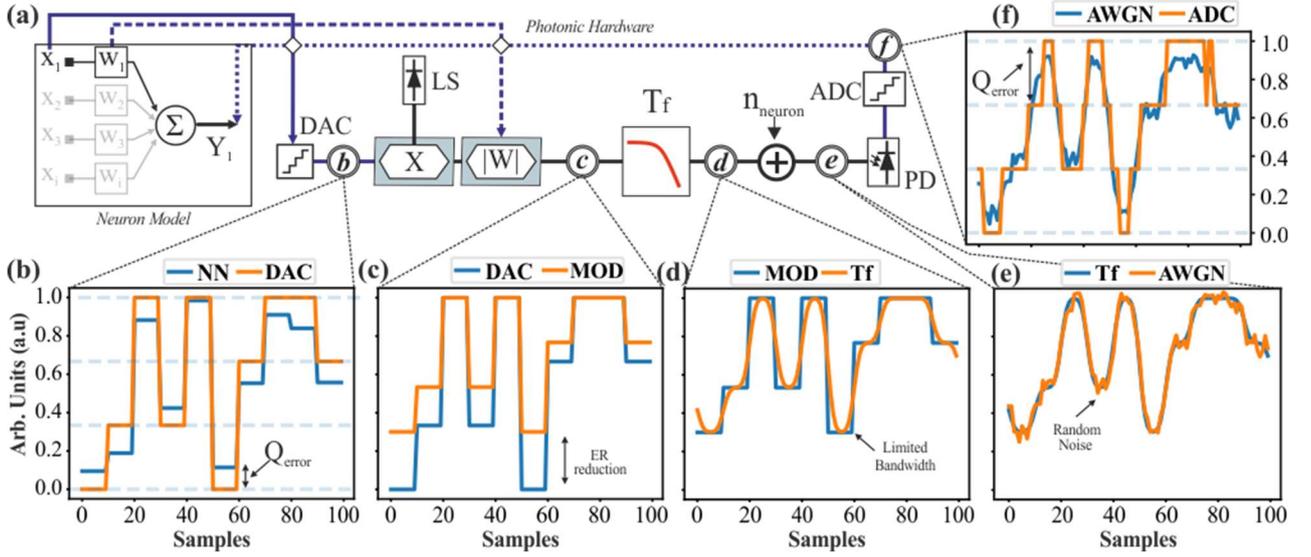

**Fig. 15**: (a) Illustrative implementation of a single neuron axon in photonic hardware, including the main signal quality degradation mechanisms. Indicative time traces are provided across the signal path: (b) NN data before and after quantization in a DAC (c) Effect of limited ER modulator on NN data (d) Effect of limited bandwidth (e) Effect of AWGN (f) Effect of Quantization at the output ADC.

This disparity is exemplary illustrated in Fig. 15 (b), with the input NN data and the DAC having a bit resolution of 8 and 2 bit, respectively, resulting into quantization errors denoted as $Q_{error}$. Followingly, the quantized electrical signal at the DAC's egress is used to drive an optical modulator in order to imprint the information in the optical domain. In this case, the non-linearity and non-infinite ER of the photonic modulator will modify the incoming signal, with Fig. 15 (c) indicatively illustrating the effect on the signal representation of limited ER. It should be pointed out, that in this simple analysis, we assume a weight stationary layout and as such neglect the effect of weight noise. We also approximate the frequency response of the photonic axon, denoted as $T_f$ as a low-pass filter, a valid assumption when considering the convolution of the constituent frequency responses of the modulator and the photodiode that are typically limited in the GHz range. The effect of this low-pass behaviour is schematically captured in Fig. 15 (d), showcasing the effect of limited bandwidth on the calculated weighted sum. Several noise-sources also degrade the SNR of the optical signal traversing the photonic neuron, including among others Relative Intensity Noise (RIN), shot noise and thermal noise. Under the general valid assumption that the main noise contribution can be approximated as AWGN sources, we concatenate the noise profile of the photonic axon into a single noise factor, correlated with the standard deviation of the zero-mean AWGN added to the signal. Fig. 15 (e) illustrates the effect of random AWGN noise added on the neural data that has propagated through the photonic hardware. Finally, an ADC is utilized for interfacing the signal back to the digital domain, introducing again quantization error, as depicted in Fig. 15 (f). Comparing the finally received digital signal at the ADC output with the original NN input digital signal can clearly indicate the significant differences that may translate into degraded NN performance when relying on conventional DL training models.

In this section, we begin by highlighting the challenges and opportunities of using photonic activation functions in NN implementations, followed by an in-depth analysis of the approach and related benefits of incorporating photonic noise, limited bandwidth, limited ER and quantization in NN training. Finally, we provide a brief overview of related applications and discuss the potential of Optics-Informed DL models.

### A. Training with Photonic Activation Functions: Challenges and Solutions

Non-linear transfer functions provided by the photonic substrates are integrated into the training process by fitting generic functions, such as sigmoid, sinusoidal, and tanh, to the experimental observation of the physical response of the components [114], [118], [119]. In turn, the fitted transfer functions are employed in software-implemented neural networks during training. More specifically, authors in [119] presented an all-optical neuron that utilizes a logistic sigmoid activation function, using a WDM input and weighting scheme. The activation function is realized by means of a deeply-saturated differentially-biased Semiconductor Optical Amplifier-Mach-Zehnder Interferometer (SOA-MZI) [133] followed by a SOA-Cross-Gain-Modulation (XGM) gate. The transfer function of the photonic sigmoid activation function is defined as:

$$g_{sig}(z) = A_2 + \frac{A_1 - A_2}{1 + e^{(z-z_0)/d}} \in \mathbb{R}, \quad (20)$$

in which the parameters $A_1 = 0.060, A_2 = 1.005, z_0 = 0.154$ and $d = 0.033$ are tuned to fit the experimental observations as implemented on real hardware devices [119].

Furthermore, the photonic sinusoidal is another activation function that is widely used in benchmarks and photonic hardware. The photonic layout corresponds to employing a MZM device [134] that converts the data into an optical signal along with a photodiode. The formula of this photonic activation function is the following:

$$g_{sin}(z) = \begin{cases} 0, & \text{if } z < 0, \\ \sin\frac{\pi^2}{2}z, & \text{if } 0 < z < 1, \\ 1, & \text{if } z > 1. \end{cases} \quad (21)$$

Recently, authors in [118] demonstrated a programmable analog opto-electronic (OE) circuit that can be configured to provide a range of nonlinear activation functions for incoherent neuromorphic photonic circuits at up to 10 GHz line rates. The proposed programmable OE circuit provides activation functions similar to those typically used in DL models including tanh-, sigmoid-, ReLU-, and inverted ReLU-like activations. Additionally, it introduces also a series of novel photonic nonlinear functions that are referred to as Rectified Sine Squared (ReSin), Sine Squared with Exponential tail (ExpSin) and Double Sine Squared.

The provided analog non-linear activation integrated circuit (NLA-IC) offers the capability to implement two different sets of activation functions. The first comprises the OE activations where the NLA-IC is exploited as a standalone activation unit, producing a variety of activation functions that are described by a generic tanh mathematical function, based on the response of the programmable transimpedance amplifier (TIA) [135], and is given by:

$$g_{nla}(x) = k\frac{a + b\sinh(x - x_0)}{c + d\cosh(x - x_0)} \in \mathbb{R}, \quad (22)$$

where the $a, b, c, d, k \in \mathbb{R}$ are hyper parameters and define the different behaviors of the programmable circuit. The second set of AFs incorporates the OEO activations. In this case, the NLA-IC is combined with a MZM producing a diverse range of OEO non-linear activation functions syndicating the OE responses and the MZM's sine-squared response. The mathematical function describing the OEO functions is given by:

$$g_{mzm}(g_{nla}(x)) = e\sin^2\left(\frac{a + b\sinh(x - x_0)}{c + d\cosh(x - x_0)}\right) + h \in \mathbb{R}, \tag{23}$$

where, as before, the $e, d \in \mathbb{R}$ defines the behavior of the programmable circuits. Such transfer functions are integrated during training to simulate the inference process on actual photonic devices. However, training models that utilize such activation functions get challenging due to the limited activation window they offer, making them susceptible to vanishing gradient phenomena [136], [137]. More precisely, such difficulties are attributed to the physical properties which force the activation to work on a smaller region of the input domain, leading to a narrow activation window and making them easily saturated. These limitations arise from the fact that physical systems operate within a specific power range, while low power consumption is also a parameter that must be taken into account during hardware design and implementation. These limitations, which are further exaggerated when recurrent architectures are used, dictate the employment of different training paradigms [63], [138]. Indeed, different activation functions require the use of different initialization schemes to ensure that the input signal will not diminish and that the gradients will correctly back-propagate. Failing to use an initialization scheme that is correctly designed for the activation function at hand can stall the training process or lead to sub-optimal results [139]. Therefore, even though these photonic neuromorphic implementations can significantly improve the inference speed, further advances are required in the way that NNs are designed and trained in order to fully exploit the potential of such photonic hardware.

Motivated by the variance preserving assumption [136], novel initialization approaches, targeting photonic activation functions, analytically compute the optimal variance during the initialization [114]. More advanced approaches propose activation agnostic methods applying auxiliary optimization tasks that allow initializing neural network parameters by taking into account the actual data distribution and the limited activation range of the employed transfer functions [140].

### B. Noise-aware Training in Optics-Informed DL

The aforementioned approaches lowered the performance gap between software and hardware implemented architectures [31]. However, in hardware-implemented neural networks, there are also limitations that arise due to the noise that emerges from various sources, such as shot noise, thermal noise, and weight read noise [31], [57], [98], as well as due to other phenomena, such as limited bandwidth and extinction ratio. To this end, there are approaches that model such noise sources as AWGN. In this way, the effect of noise is introduced in the software training process, significantly improving the performance during the deployment, exploiting the robustness of ANNs to noise, especially when they take it into account during the training process [31]. More specifically, noise sources are simulated in order to train in a noise-aware fashion. In this way, we exploit the fact that DL models are intrinsically robust to noise, especially when they are first adequately trained to tolerate noise sources, which are modeled as AWGN. Therefore, the output of a neuron that incorporates such a source can be modeled as:

$$u_{ti}^{(r)} = \sum_{j=1}^{M} w_{ij}^{(in)}\left(x_{tj} + \mathcal{N}(0, \sigma_i^2)\right) + \mathbf{w}_i^{(r)^T} \mathbf{y}_{t-1}^{(r)} + \mathcal{N}(0, \sigma_w^2), \qquad (24)$$

where $\sigma_i^2$ and $\sigma_w^2$ are the variance of the noise that affect the input and weighting, respectively. Also, the final output of the neuron is modeled as:

$$y_{ti}^{(r)} = f(u_{ti}^{(r)}) + \mathcal{N}(0, \sigma_a^2), \qquad (25)$$

where $\sigma_a^2$ is the variance of the noise induced by the activation function. Training DL models using the aforementioned noise sources makes them more robust to noise that will arise during deployment.

On top of that, the authors in [141] propose an advanced initialization method that incorporates the existing noise sources during the initialization method, estimating iteratively the optimal variance for each layer by taking into account the actual data distribution and as such, leading to significant performance improvements during the deployment. To this end, in [141] the authors propose an auxiliary task-based approach that can estimate the initialization variance for each layer. To this end, they introduce an additional scale factor $a_i$ for each layer,

$$y^{(i)} = f(|a_i| \mathbf{W}_i^{\top} \mathbf{y}^{(i-1)} + \mathbf{b}_i) \qquad (26)$$

where $|\cdot|$ denotes the absolute value operator. Assuming that the weights are initialized by drawing from a Gaussian distribution with zero mean and variance denoted by $N(0, \sigma^2)$, altering the scaling factors, results in adjusting the initialization variance for each layer. Moreover, in order to optimize the scaling factor $a_i$ an auxiliary linear classification layer is required, $W_i^{class} \in \mathbb{R}^{m^{(i)} \times N_C}$, where $N_C$ is the number of classes (for a multi-class classification task) or the number of values to regress (for regression problem). In this way, $a_i$ and $W_i^{class}$ are those terms that need to be optimized, while the actual weights and biases of the network are kept fixed. Then, the output of the classification layer can be directly used to perform the task at hand, i.e., either classification or regression. Additionally, an extra regularization penalty parameter is added, denoted by $\Omega(a_i)$, in an effort to penalize scaling factors that lead to saturation of the activation function. Specifically, after forward passing from linear term, $z_i = |a_i|W_i x + b_i$, we calculate $\Omega(a_i)$ as following:

$$\Omega(a_i) = \frac{\max\{l - z_i, z_i - u, 0\}}{nm}, \tag{27}$$

where $l$ and $u$ is the lower and upper bound of activation region, while $n$ and $m$ is the fan-in and fan-out respectively and $\max\{\cdot\}$ denotes the maximum element in the set. $\tilde{J}(W_i^{class}, a_i; X, y)$ is formulated as:

$$\tilde{J}(W_i^{class}, a_i; X, y) = J(W_i^{class}, a_i; X, y) + c\Omega(a_i). \tag{28}$$

Finally, the scaling factor $a_i$ and classification weights layers are optimized using gradient descent. After the optimization has been completed, the weights of the $i$-th layer can be re-initialized using the optimized scaling factor $a_i$. All layers of the network, from input to output, are iteratively initialized with the aforementioned procedure. This initialization scheme considers the modeled noise sources and can appropriately adjust the variances accordingly. After this process has been completed, the model is ready to be trained using regular back-propagation. The ability of neural networks to compensate for such phenomena by taking them into account during the training have been also used to decompose different noise source and introduce them during the training process. Indeed, such approaches have been also used successfully to handle the limited bandwidth and extinction ratio [130].

*C. Training with Quantized and Mixed Precision Neural Networks*

Other operations, such as ADC and DAC operations, have also been shown to affect the accuracy of photonic neural networks. However, when these phenomena are considered during the training, results in more robust representations and, in turn, in higher performance during the deployment. More specifically, photonic computing includes the employment of DAC and ADC conversions along with the parameters encoding, amplification, and processing devices, such as modulators, PDs and amplifiers, which, inevitably, introduce degradation of the analog precision during inference, as each constituent introduces a relevant noise source that impacts the electro-optic link's bit resolution properties. Thus, the noise introduced increases when higher line rates are applied, translating to lower bit resolution. Furthermore, being able to operate in lower precision networks during deployment can further improve the potential use of analog computing by increasing the computational rate of the developed accelerators, while keeping energy consumption low [53], [142].

Typically, the degradation introduced to analog precision can be simulated through a quantization process that converts a continuous signal to a discrete one by mapping its continuous set to a finite set of discrete values. This can be achieved by rounding and truncating the values of the input signal. Despite the fact that quantization techniques are widely studied by the DL community [142]-[144], they generally target large CNNs containing a large number of surplus parameters with a minor contribution to the overall performance of the model [145], [146]. Furthermore, existing works in the DL community focus mainly on partially quantized models that ignore input and bias [143], [147]. These limitations, which are further exaggerated when high-slope photonic activations are used, dictate the use of different training paradigms that take into account the actual physical implementation [31]. Indeed, neuromorphic photonics impose new challenges on the quantization of the DL model, requiring the appropriate adaptation of existing methodologies to the unique limitations of photonic substrates. Furthermore, the quantization scheme applied in neuromorphic photonics typically follows a very simple uniform quantization [57], [148]. This differs from the approaches traditionally used in trainable quantization schemes for DL models [149], as well as mixed precision quantization [150].

To this end, several proposed approaches deal with the limited precision requirements before models are deployed to hardware. Such approaches calibrate networks with limited precision requirements after training the models, name post-training quantization methods, offering improvements in contrast to applying the model directly to hardware without taking into account the limited precision components [150]. Other approaches take into account the limited precision requirements during training, naming quantization-aware training methods [150], [151]. Later methods have shown that significantly exceed the performance of post-training approaches, eliminating or restricting performance degradation between the full precision and limited precision models [151].

The authors in [131] proposed an activation-agnostic, photonic-compliant, and quantization-aware training framework that does not require additional modifications of the hardware during inference, significantly improving model performance at lower bit resolution. More specifically, they proposed to train the networks with quantized parameters by applying uniform quantization to all parameters involved during the forward pass and, consequently, the quantization error is accumulated and propagated through the network to the output and affects the employed loss function. In this way, the network is adjusted to lower-precision signals, making it more robust to reduced bit resolution during inference, significantly improving model performance. To this end, every signal involved in the response of the $i$-th layer is first quantized in a specific floating range $\left[h_{min}^{(i)}, \dots, h_{max}^{(i)}\right) \in R$. Then, during the forward pass of the network, quantization error $\epsilon$ is injected to simulate the effect of rounding during the quantization, while during the backpropagation the rounding is ignored and approximated with an identity function. A comprehensive mathematical analysis regarding the quantization training can be found in [150], [151]. Finally, more advanced approaches targeting novel dynamic precision architectures [107], [152], propose stochastic approaches to gradually reduce the precision of layers within a model, exploiting their position and tolerance to noise, based on theoretical indications and empirical evidence [153]. More specifically, the stochastic mixed precision quantization-aware training scheme, which is proposed in [153], adjusts the bit resolutions among layers in a mixed precision manner, based on the observed bit resolution distribution of the applied architectures and configurations. In this way, the authors are able to significantly reduce the inference execution times of the deployed NN [107].

*D. Applications*

Applying the aforementioned methods allows us to employ PNNs where high frequencies with minimum energy consumption are required, utilizing the DL techniques in a whole new spectrum of applications. Such applications include network monitoring and optical signal transmission, where the high compute rates limit the application of existing accelerators. For example, neuromorphic photonics are capable of operating at very high frequencies and can be integrated on a backplane pipeline of a modern high-end switch, which makes them an excellent choice for challenging Distributed Denial of Service (DDoS) attacks detection applications, where high-speed and low-energy inference is required. More specifically, the authors in [37], [154] build on the concept of a neuromorphic lookaside accelerator, targeting to perform real-time traffic inspection, searching for DDoS attack patterns during the reconnaissance attack phase, when the attacker tries to determine critical information about the target's configuration. Before deploying a DDoS attack, a port scanning procedure is compiled to track open ports on a target machine. During this procedure, port scanning tools, such as Nmap, create synthetic traffic that can be captured and analyzed by the proposed network, capturing huge amounts of packages in the used computation rates of the modern high-end switches.

Another domain that can be potentially benefited from neuromorphic hardware is communications. Over the recent years, there is an increasing interest in employing DL in the communication domain [155], ranging from wireless [156] to optical fiber communications [42], exploiting the robustness of ANNs to noise, especially when they take it into account during the training process. Such approaches design the communication system by carrying out the optimization in a single end-to-end process, including the transmitter, receiver, and communication channel, with the ultimate goal to achieve optimal end-to-end performance by acquiring a robust representation of the input message [42], [157] introduced an end-to-end deep learning fiber communication transceiver design, emphasizing training by examining all optical activation schemes and respective limitations present in realistic demonstrations. They applied the data-driven noise-aware initialization method [158] that is capable of initializing PNNs by taking into account the actual data distribution, noise sources as well as the unique nature of photonic activation functions. They focused on training photonic architectures which employ all optical activation schemes [119], by simulating their given transfer functions. This allows for reducing the effect of vanishing gradient phenomena, as well as improving the ability of networks coupled with communication systems to withstand noise, e.g., due to the optical transmission link. As experimentally demonstrated, this method is significantly tolerant to the degradation occurred when easily saturated photonic activations are employed as well as significantly improves the signal reconstruction of the all-optical intensity modulation/direct detection (IM/DD) system.

**Conclusion**

Conventional electronic computing architectures face many challenges due to the rapid growth of compute power, driven by the rise of AI and DNNs, calling for a new hardware computing paradigm that could overcome these limitations and being capable of sustaining this ceaseless compute expansion. In this tutorial, prompted by the ever-increasing maturity of silicon photonics, we presented the feasibility of PNNs and their potential embodiment in future DL environments. First, we discussed the essential concepts and criteria for NN hardware, examining the fundamental components of NNs and their core mathematical operations. Then, we investigated the interpendence of analog bit precision and energy efficiency of photonic circuits, highlighting the benefits and challenges of PNNs over conventional approaches. Moreover, we reviewed the state-of-the-art PNN architectures, analyzing their perspectives with respect to MVM operations execution, weight technology selection and activation function implementation. Finally, the recently introduced optics-informed DL training framework was presented, which comprises a novel software-hardware NN co-design approach that aims to significantly improve the NN accuracy performance by incorporating the photonic hardware idiosyncrasy into NN training.

*Appendix: Digital-analog-precision loss*

In order to get a meaningful sense of the relationship between digital-to-analog precision loss and input optical power in a photonic matrix multiplier, we begin by assuming a linear PAM-M optical signal, or equivalently a signal with a bit resolution $B = \log_2 M$, featuring infinite ER, and an average optical power of $P_{avg}$. When receiving the signal in a thermal noise dominated optical link, we can evaluate its quality using the Q factor of the outer eye diagram of PAM-M modulation, that can be expressed through:

$$Q_{outer-eye} = \frac{P_1 - P_m}{\sigma_1 + \sigma_m} = \frac{P_1 - P_m}{2 \times \sigma_t} \tag{A1}$$

, where $P_1$ is the optical signal's peak power, $P_m$ the optical power of the signal's penultimate level and $\sigma_t$ the standard deviation of the thermal noise. The optical power of the penultimate level of a linear PAM-M signal can be calculated through subtracting the distance between the penultimate PAM-M level from the peak power:

$$P_m = P_1 - \frac{P_1 - P_0}{2^B - 1} = P_1 - \frac{P_1}{2^B - 1} \tag{A2}$$

, with $P_0 = 0$ for an infinite ER signal. Replacing (A2) into (A1):

$$Q_{outer-eye} = \frac{P_1 - P_m}{\sigma_1 + \sigma_m} = \frac{P_1 - P_1 + \frac{P_1}{2^B - 1}}{2 \times \sigma_t}$$

$$Q_{outer-ey} = \frac{1}{2^B - 1} \times \frac{P_1}{2 \times \sigma_t} \tag{A3}$$

From equation (A3) we can deduct that for a given bit resolution $B = \log_2 M$, in order to maintain the same signal quality we need to increase the optical power of the receiver's input signal.

In the case of the analog matrix multiplier, assuming a loss-less weight matrix implementation, $N$ inputs signals of (PAM-M) and only positive weight values the optical peak power of the signal emerging at the output can be calculated through:

$$P_{out} = \sum_{i=1}^{N} P_i \quad (4) \tag{A4}$$

, where $P_i$ is the optical peak power of each constituent signal. When the signals have the same optical peak power $P_i$ we can transform (A4) to:

$$P_{out} = N \times P_i \tag{A5}$$

Moreover, for an $N \times N$ optical matrix multiplier, the input optical signal will experience loss from the front-end optical splitter, that can be calculated from:

$$S_{loss} = N + S_{e\_loss} \tag{A6}$$

, where $S_{e\_loss}$ is the excess loss of the splitter, that for a cascaded tree MMI layout can be calculated through $S_{e\_loss} = MMI_{loss} \times log_2 N$. As such equation (A5) can be rewritten as:

$$P_{out} = \frac{N \times P_i}{log_2 N \times MMI_{loss} + N} \tag{A7}$$

Assuming a Silicon Photonic implementation and some state-of-the-art values for MMI-loss i.e. $MMI_{loss} = 0.06\ dB$ or $MMI_{loss} = 0.014$ in natural numbers [159] we can deduct that the aforementioned equation can be simplified without significant accuracy loss in:

$$P_{out} = \frac{N \times P_i}{N} \tag{A8}$$

as $log_2 N \times 0.014 \ll N$.

Based on the above we will highlight two distinct cases. In the first that we refer to as full digital precision $a_{prec} = 1$, we increase the optical power of the input laser by a factor of the beam splitter loss i.e. $N$, essentially compensating the optical loss of the splitter for each contributing beam as such:

$$P_{laser\_a\_1} = \frac{N}{a_{prec}} \times P_{laser} = N \times P_{laser}$$

$$P_{out\_a\_1} = \frac{N \times P_i}{N} \times \frac{N}{a_{prec}} = N \times P_i \tag{A9}$$

This safeguards that the minimum optical power difference between adjacent bits (MOPB) remains constant such as:

$$MOPB_{P1} = MOPB_{P_{out\_a\_1}}$$

$$P_i - P_m = P_{out\_a\_1} - P_{M-1'}$$

$$\frac{1}{2^B - 1} \times P_i = \frac{1}{2^{B^{out}} - 1} \times P_{out\_a\_1}$$

$$\frac{1}{2^B - 1} \times P_i = \frac{1}{2^{B^{out}} - 1} \times N \times P_i$$

$$\frac{1}{2^B - 1} = \frac{1}{2^{B^{out}} - 1} \times N$$

$$B^{out} = log_2(N \times 2^B - N + 1) = log_2(N \times (2^B - 1) + 1) \tag{A10}$$

Using exemplary values from the above equation we can see that the equivalent bit resolution of the output signal is significantly higher than the input signal e.g. for N=4 and M=2, $M^{out} = 3.7$.

Another interesting operational regime, defined as $a_{prec} = N$, where we keep the output equivalent bit precision $B^{out}$ the same as the input signal resolution $B$, such as $B^{out} = B$. In this case the signal quality at the output defined through the Q factor remains the same as the input $P_i$:

$$Q_{outer-eye_{P_i}} = Q_{outer-ey\ P_{out}}$$

$$\frac{1}{2^B - 1} \times \frac{P_i}{2 \times \sigma_t} = \frac{1}{2^{B^{out}} - 1} \times \frac{P_{out}}{2 \times \sigma_t}$$

$$P_i = P_{out} \tag{A11}$$

From the above equation we can deduce that when we maintain the same equivalent bit resolution at the output of the matrix multiplier, we don't need to compensate for the splitter loss as we are effectively trading the bit resolution for reduced laser power. Comparing to the full digital precision case we have:

$$P_{laser\_a\_N} = P_{laser} = \frac{P_{laser\_a\_1}}{N} \tag{A12}$$

, where $P_{las\_a_N}$, $P_{laser_{a_1}}$ are the required optical powers for full digital precision ($a_{prec} = 1$) and same input-output bit resolution ($a_{prec} = N$)

Summing up, we defined $a_{prec}$ as the analog-digital precision and illustrated two operational regimes:

- $a_{prec} = 1$, when we increase the output power of the laser source to compensate for the splitting loss by a factor of N. In this case the output bit resolution reaches $B^{out} = (N \times (2^M - 1) + 1)$, that for only integer values of bit resolution can be simplified to $B^{out} = B + log_2(N)$
- $a_{prec} = N$, where we keep the same bit precision at both the input and the output, trading off the decreased bit precision, as opposed to the full digital precision case, with lower input laser power by a factor of $a_{prec} = N$

**Acknowledgements**

The work was in part funded by the EU Horizon projects PlasmoniAC (871391), SIPHO-G (101017194) and Gatepost (101120938).

13π tuning range," Opt. Express 29(4), 5525-5537 (2021).

[103].     N. Quack, H. Sattari, A. Y. Takabayashi, Y. Zhang, P. Verheyen, W. Bogaerts, P. Edinger, C. Errando-Herranz and K. B. Gylfason, "MEMS-Enabled Silicon Photonic Integrated Devices and Circuits," IEEE Journal of Quantum Electronics, vol. 56, p. 1–10, February 2020.

[104].     R. Amin, R. Maiti, Y. Gui, C. Suer, M. Miscuglio, E. Heidari, R. T. Chen, H. Dalir, V. J. Sorger, "Sub-wavelength GHz-fast broadband ITO Mach–Zehnder modulator on silicon photonics," Optica 7(4), 333-335 (2020).

[105].     Sorianello, V., Midrio, M., Contestabile, G. et. al., Graphene–silicon phase modulators with gigahertz bandwidth. Nature Photon 12, 40–44 (2018).

[106].     William M. J. Green, Michael J. Rooks, Lidija Sekaric, and Yurii A. Vlasov, "Ultra-compact, low RF power, 10 Gb/s silicon Mach-Zehnder modulator," Opt. Express 15, 17106-17113 (2007).

[107].     G. Giamougiannis, et. al., "Analog nanophotonic computing going practical: silicon photonic deep learning engines for tiled optical matrix multiplication with dynamic precision" Nanophotonics, vol. 12, no. 5, 2023.

[108].     K. Kawaguchi, "Deep learning without poor local minima," in Advances in Neural Information Processing Systems, (2016), pp. 586–594

[109].     X. Glorot, A. Bordes, and Y. Bengio, "Deep Sparse Rectifier Neural Networks," (2011), pp. 315–323

[110].     K. He, X. Zhang, S. Ren, and J. Sun, "Delving deep into rectifiers: Surpassing human-level performance on imagenet classification," in Proceedings of the IEEE international conference on computer vision, (2015), pp. 1026–1034.

[111].     S. Hochreiter and J. Schmidhuber, "Long short-term memory," Neural computation 9, 1735–1780 (1997).

[112]. J. K. George, H. Nejadriahi, and V. J. Sorger, "Towards on-chip optical ffts for convolutional neural networks," in 2017 IEEE International Conference on Rebooting Computing (ICRC),